\documentclass[pdflscape,lscape,iop,apj,numberedappendix,appendixfloats,stfloats,url,microtype,iop,revtex4]{emulateapj}

\usepackage{paralist,amsmath,amssymb}
\usepackage{longtable,rotating}
\usepackage{float}
\usepackage{morefloats}
\usepackage{placeins}
\usepackage{xfrac}
\usepackage{pdflscape}

    {\pagebreak[4]\global\pdfpageattr\expandafter{\the\pdfpageattr/Rotate 90}}%
    {\pagebreak[4]\global\pdfpageattr\expandafter{\the\pdfpageattr/Rotate 0}}%

\usepackage{threeparttable}
\shorttitle{Low-polarization patches in Fornax A}
\shortauthors{Anderson et al.}

\begin{document}


\title{Broadband radio polarimetry of Fornax A, I: Depolarized patches generated by advected thermal material from NGC 1316}


\author{
C. S. Anderson\altaffilmark{1,2}$^{\dagger}$,
B. M. Gaensler\altaffilmark{2,3},
G. H. Heald\altaffilmark{1},
S. P. O'Sullivan\altaffilmark{4},
J. F. Kaczmarek\altaffilmark{2},
I. J. Feain\altaffilmark{5}
}

\altaffiltext{$\dagger$}{{\bf craig.anderson@csiro.au}}
\altaffiltext{1}{CSIRO Astronomy \& Space Science, Kensington, Perth 6151, Australia}
\altaffiltext{2}{Sydney Institute for Astronomy (SIfA), School of Physics, University of Sydney, NSW 2006, Australia}
\altaffiltext{3}{Dunlap Institute for Astronomy \& Astrophysics, University of Toronto, Ontario, Canada}
\altaffiltext{4}{Hamburger Sternwarte, Universit\"{a}t Hamburg, Gojenbergsweg 112, 21029 Hamburg, Germany}
\altaffiltext{5}{Central Clinical School, School of Medicine, University of Sydney, NSW 2006, Australia}


\begin{abstract}

We present observations and analysis of the polarized radio emission from the nearby radio galaxy Fornax A over 1.28--3.1 GHz, using data from the Australia Telescope Compact Array (ATCA). In this, the first of two associated papers, we use modern broadband polarimetric techniques to examine the nature and origin of conspicuous low-polarization (low-$p$) patches in the lobes. We resolve the low-$p$ patches, and find that their low fractional polarization is associated with complicated frequency-dependent interference in the polarized signal generated by Faraday effects along the line of sight. The low-$p$ patches are spatially correlated with interfaces in the magnetic structure of the lobe, across which the line-of-sight-projected magnetic field changes direction. Spatial correlations with the sky-projected magnetic field orientation and structure in total intensity are also identified and discussed. We argue that the low-$p$ patches, along with associated reversals in the line-of-sight magnetic field and other related phenomena, are best explained by the presence of $\mathcal{O}(10^9)$ $M_\odot$ of magnetized thermal plasma in the lobes, structured in shells or filaments, and likely advected from the ISM of NCG 1316 or its surrounding ICM. Our study underscores the power and utility of spatially-resolved, broadband, full-polarization radio observations to reveal new facets of flow behaviors and magneto-ionic structure in radio lobes and their interplay with the surrounding environment. 

\end{abstract}


\section{Introduction}\label{sec-intro}


The radio galaxy Fornax A provides an excellent opportunity to study how radio lobes interact with their external environment. The system has undergone a series of dynamical interactions (e.g. \citealp{Schweizer1980,Horellou2001,Lanz2010,Beletsky2011,Iodice2017}) that have triggered episodic activity in the central active galactic nucleus (AGN; \citealp{Lanz2010} and references therein) leading to the formation of the iconic radio lobes. However, despite the presence of a weak radio jet in the core of the host galaxy NGC 1316 \citep{Geldzahler1984}, the system appears to have evolved relatively passively over the past 0.1 Gyr \citep{Iyomoto1998,KF2003}. The lobes themselves are large and bright, do not have hotspots, do not exhibit the pronounced edge darkening or brightening characteristics of Fanaroff-Riley (FR) I or II sources, and are not connected to the radio core by visible jets. They are quite old (0.1--1 Gyr; \citealp{Ekers1983}; see also Section \ref{sec-summaryofproperties}), and their expansion is either frustrated or confined by the relatively sparse Fornax intracluster medium (ICM; \citealp{Drinkwater2001,Paolillo2002,Scharf2005,Seta2013}). In summary, the Fornax A radio lobes are currently in a relatively mature stage of their evolution, presumably after having interacted more vigorously with the external ICM, and a more spatially-limited ISM \citep{KFM1998,KF2003}, in previous epochs.

The lobes host a pronounced network of patches or filaments distinguished by their low apparent emissivity in linear polarization \citep{Fomalont1989}. These structures form the subject of this paper, and we discuss them further below. However we first define some terminology, for the system plays host to a variety of small- to medium-scale structure, in both total and polarized emission intensity \citep{Fomalont1989}. We use the term ``low-$p$ patches" to collectively refer to patchy/filamentary morphological features that are distinguished by their low level of both fractional and absolute linearly polarized intensity. We use ``Stokes $I$ filaments'' and ``polarized filaments'' to refer to quasi-linear structures that have high total and polarized emission intensities (respectively) relative to their surroundings. Where we use the word ``depolarize'' and its derivatives without further qualification or modifying context, we are referring to Faraday depolarization (e.g. see \citealp{GW1966}). All of these morphological features are described further in Section \ref{results}. 

The physical cause of the low-$p$ patches in Fornax A remains debatable. In other sources, similar structures have been attributed to interactions between the lobe and environment, invoking processes such as shocks \citep{Perley1984}, surface waves \citep{Mason1988,Bicknell1990}, the formation and evolution of jet-blown bubbles \citep{vBF1984}, cooling instabilities \citep{SA1967,dGPO1989}, an exterior medium \citep{Clarke1992}, and hydrodynamic instabilities in the magnetized fluid (e.g. \citealp{Bicknell1990}). In addition, large and complicated Faraday rotation measure (RM) fluctuations across nearby radio lobes have long been observed (e.g. \citealp{Dreher1987,Owen1990}), and these may also be related to low-$p$/depolarized patches. While these RM structures are often attributed to the foreground ICM, in some cases the RM structure appears to be generated at the interface between lobes and their surroundings. For example, \citet{Carilli1988} argue that the RM enhancements detected towards the hotspots of Cygnus A are generated as a bow shock is driven into the ICM by the associated radio jet. \citet{VT1999} argue that RM enhancements observed in the vicinity of the radio jet in 3C 216 arize as a result of jet-ISM interaction in the host galaxy. More recently, \citet{Guidetti2011,Guidetti2012} have argued that large-scale, ordered, linear RM structures observed towards several lobed sources are generated at or near the locus of interaction with their surrounding environment. 

The past $\sim$decade has also seen important progress in semi-analytic descriptions, and in hydrodynamic and magnetohydrodynamic simulations, of radio lobes in various environments. These studies (and their forerunners) have uncovered a variety of mechanisms through which low-polarization or depolarized filamentary structure might be generated in and around radio lobes. For example, buoyantly advected thermal plasma from the host galaxy could act as a Faraday-depolarising screen for emission lying in the background (e.g. \citealp{Churazov2001,Reynolds2002,Pope2010}). Alternatively, lobes are generally found to be subject to various hydrodynamical or magneto-hydrodynamical instabilities that generate a complex vortical flow behavior resulting in complex magnetized structure, both internally and at their boundaries (e.g. \citealp{Bicknell1990,BK2001,Kaiser2005,Reynolds2005,KA2007,DF2008,DS2009,HE2011,Roediger2013,HK2014,TS2015,EHK2016}; and refs. in each). Associated depolarization might then be generated, for example, by foreground turbulence, mixed synchrotron-emitting and Faraday rotating plasmas, or crossed magnetic field lines.

In regards to Fornax A itself, \citet{Fomalont1989} suggested that thermal plasma in the lobes could generate the low-$p$ patches through Faraday depolarization, consistent with the buoyant advection scenario described in the preceding paragraph. It is an enticing hypothesis: Detection of thermal plasma in radio lobes has long been sought, since, in addition to encoding information about the interaction history of the system, such material can contribute significantly to the pressure balance, dynamics, and energetics of radio lobes, and can act as seed material for high energy cosmic rays (e.g. \citealp{Hardcastle2010}). Whilst early searches for such material did not result in direct detections (\citealp{MN2007} and references therein), more recent studies at radio, X-ray, and $\gamma$-ray wavelengths have claimed as much in the lobes of our two nearest radio galaxies --- Fornax A \citep{Seta2013,McKinley2015,Ackermann2016} and Centaurus A \citep{OSullivan2013b,Stawarz2013} --- though the authors were not able to place strong constraints on the spatial distribution and magneto-ionized structure of this material.  

Thus, hypothesising that the low-$p$ patches are likely to provide important insights into the nature, physics, and timeline of dynamical processes operating in and around the lobes, we have undertaken a broadband polarimetric analysis of these structures. Our aims are to better constrain their physical origin, and to consider the attendant implications for the magneto-ionized structure of the lobes, and for their history of interaction with the external environment. Our analysis of the main body of the lobes (i.e. outside the low-$p$ patches) will be presented in a separate paper (hereafter referred to as paper 2). This paper is organized as follows. We describe our observations and their calibration in section \ref{obscal}, followed by explanations of our imaging procedure in section \ref{imaging} and our approach to spectropolarimetric analysis in section \ref{analysis}. Our results are presented in section \ref{results}, then discussed in section \ref{discussion}. A summary of our work and conclusions are presented in section \ref{conclusion}. We assume that $H_0=70$ km s$^{-1}$ and take the distance to Fornax A to be 18 Mpc (\citealp{Feldmeier2007,Stritzinger2010}; but see \citealp{Cantiello2013}), meaning that an angular diameter of one arcminute corresponds to a linear extent of 5.2 kpc.

\section{Observations and calibration}\label{obscal}

\begin{figure*}
\centering
\hspace{-1.3cm}
\includegraphics[width=0.90\textwidth]{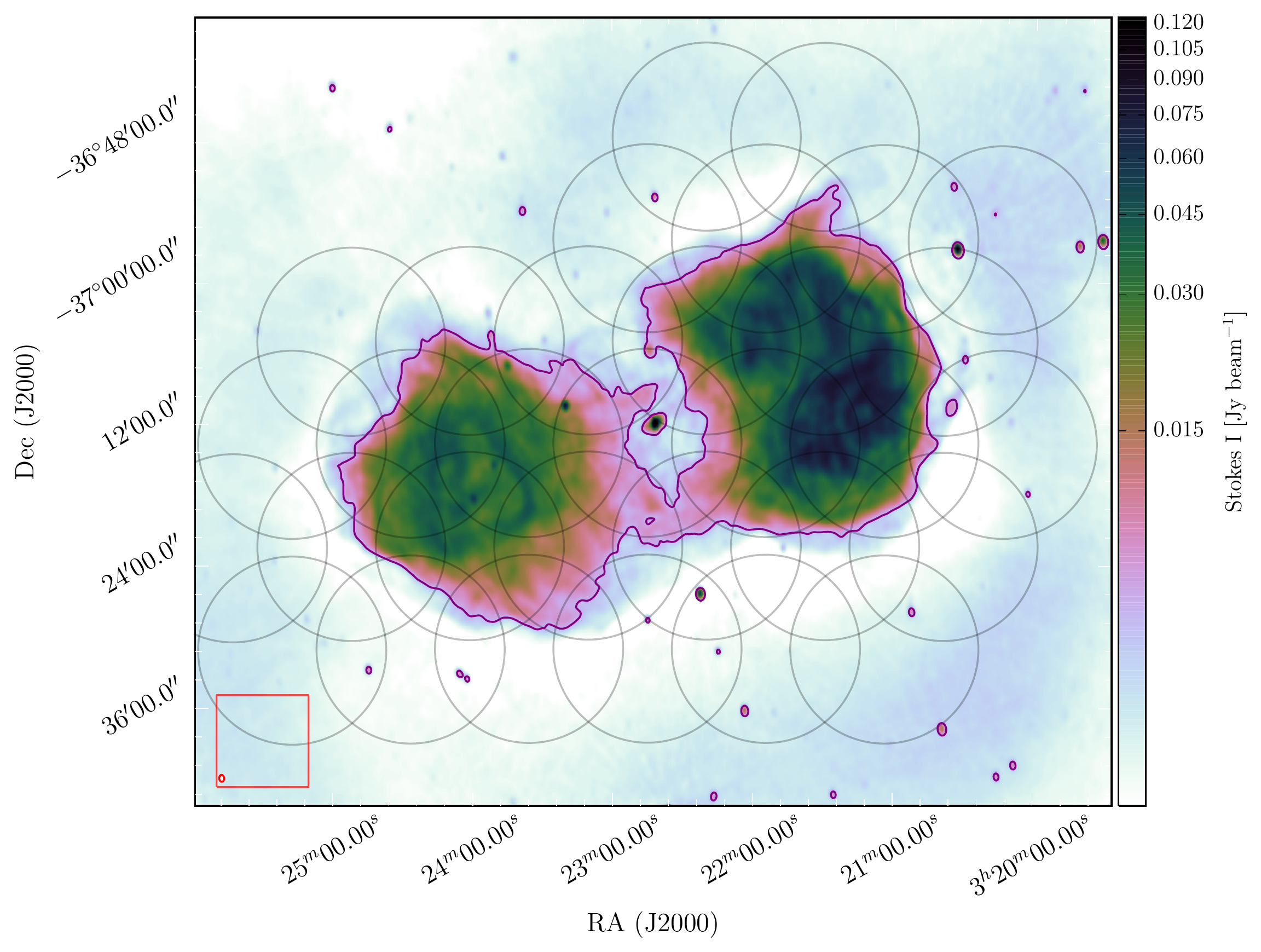}
\caption[Total intensity image of Fornax A]{An MFS Stokes $I$ map of Fornax A ($20"\times30"$ spatial resolution in RA and Decl. respectively), created as described in Section \ref{imaging}. The map employs the cubehelix colormap \citep{Green2011} applied to the data with an arcsinh stretch to emphasize structure in faint emission. The purple contour is at a flux density level of 8 mJy --- the cutoff below which we do not perform spectropolarimetric analysis. The gray circles indicate the half-power point of the ATCA primary beam at 3.1 GHz for the 32 mosaic pointings comprising our mosaic. The noise level (root-mean-square) in the red rectangle in the left-hand corner of the image is $2.4\times10^{-4}$ Jy beam$^{-1}$, while the faintest sources visible have flux densities on the order of 1 mJy beam$^{-1}$. The synthesized beam size is indicated in the bottom left-hand corner as a red ellipse.}
\label{fig:FornA_zoom}
\end{figure*}

We used the Compact Array Broadband Backend (CABB) correlator on the Australia Telescope Compact Array (ATCA; \citealp{Wilson2011}) in the `CFB 1M' mode (i.e. 2 GHz total bandwidth, 1 MHz channelization of the continuum) to observe Fornax A in full polarization between 1.1 and 3.1 GHz. We used a 32 pointing mosaic to cover the entire radio galaxy, centered on RA = $03^{h}22^{m}45^{s}$ and Decl. = $-37^{d}14^{m}00^{s}$, and laid out in a regular hexagonal grid with 0.179$^\circ$ spacing (see Figure \ref{fig:FornA_zoom}). The pointing centers of each beam fall at the one-third power point of adjacent beams at 3.1 GHz. Strictly speaking, the spatial Nyquist-sampling criterion requires adjacent pointing centers to fall at the half-power point, which maximizes sensitivity to highly extended emission. Our mosaic is critically Nyquist-sampled at a frequency of $\sim$2.8 GHz instead. This slight under-sampling at the upper end of our frequency band results in only minor non-uniformities in the image sensitivity across the field at these frequencies, and does not significantly increase the net off-axis polarization leakage in the final mosaic, nor substantially limit our ability to detect large scale structure (see further discussion in Section \ref{imaging}).  

\thispagestyle{empty}

\begin{figure}
\centering
\includegraphics[width=0.5\textwidth]{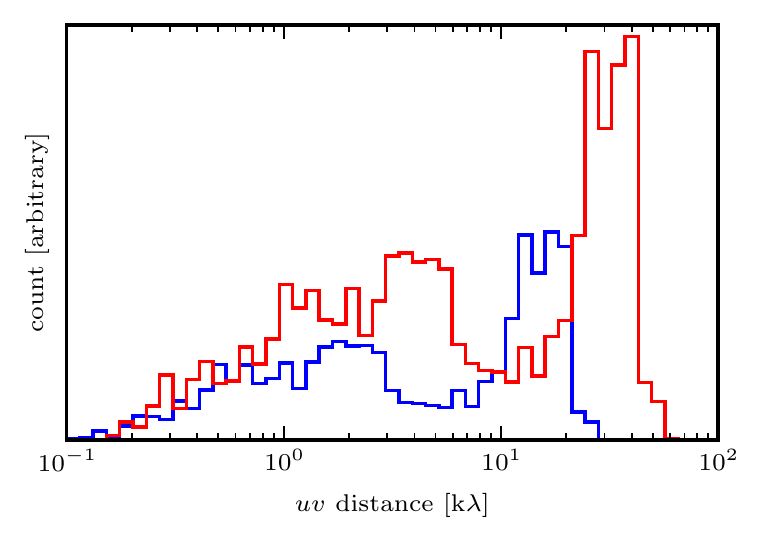}
\caption[Histogram of the $uv$ distance for data our Fornax A observations]{Histogram of the $uv$ distance for data in our Fornax A observations (in arbitrary units, with a logarithmic $x$-axis). The red (blue) histograms were generated from the $uv$ distances of visibility measurements in 200 MHz bands centered on 2.8 GHz (1.4 GHz), i.e. towards the upper (lower) end of our frequency band. Owing to the large volume of data in our measurement set, only every 13th visibility measurement was used to generate this plot.}
\label{fig:C2735_uvcoverage}
\end{figure}

We utilized each of the H75, EW352, 750B, and 6A array configurations to achieve broad $uv$ coverage at all frequencies (see Figure \ref{fig:C2735_uvcoverage}). We observed the mosaic for three separate twelve hour tracks in each of the different east-west array configurations, conducted 2012 November 3rd--5th (750B), 2013 January 11th--13th (EW352), and 2013 March 1st--3rd (6A). An additional six hour track was conducted in the H75 hybrid array, which utilizes antennas on both north-south and east-west tracks, on 2016 August 7th. Excluding overheads and array downtime, each pointing received a total integration time of $\sim$130 minutes split over a total of $\sim$130 $uv$ cuts. 

Our initial calibration followed standard procedures for the reduction of cm-band ATCA data. Daily observations of PKS B1934-638 were used to calibrate the bandpass response and absolute flux scale. The time-dependent complex antenna gains and on-axis polarization leakage were calibrated using hourly observations of PKS B0332-403. In both cases, independent calibration solutions were derived at 128 MHz intervals through the band, then interpolated across the full band, in order to account for frequency dependent variations in complex gain and polarization leakage \citep{Schnitzeler2011}. Radio-frequency interference (RFI) was flagged iteratively throughout the calibration process using the \textsc{sumthreshold} algorithm \citep{Offringa2010}. All data were flagged between 1.50 and 1.62 GHz, and below 1.28 GHz.

Following the initial calibration, we carried out three cycles of phase self-calibration, as follows. The entire mosaic was imaged jointly through the entire frequency band in 128 MHz chunks using multifrequency synthesis (MFS). We deconvolved these images using the {\sc miriad} task \textsc{pmosmem}, then convolved the maximum entropy model with a restoring beam. Then, for each imaged frequency band, we split the restored mosaics into primary-beam-weighted images of the individual mosaic pointings using the \textsc{miriad} task \textsc{demos}, then phase-self-calibrated the visibilities corresponding to that pointing and frequency range. For each mosaic pointing, the self-calibration solutions were interpolated across each of the sixteen 128 MHz sub bands spanning the full 1.28--3.1 GHz band, then applied continuously across the total bandwidth. This entire imaging and self-calibration procedure was then repeated twice more. 
 
We estimate that the post-calibration polarization leakage of Stokes $I$ into Stokes $Q$, $U$, and $V$ is no greater than 0.1\% of Stokes $I$ (e.g. \citealp{Schnitzeler2011}), and is significantly better directly on-axis when averaged over the band (see e.g. \citealp{Anderson2015,Anderson2016}) by applying RM synthesis (described below).

\section{Imaging}\label{imaging}

For our spectropolarimetric analysis, we generated Stokes $I$, $Q$, $U$, and $V$ images of Fornax A throughout the 1.28--3.10 GHz ATCA band. The data were imaged separately in 150 contiguous channels, each covering an equal range in $\lambda^2$ space of $\sim3\times10^{-4}$ m$^2$, spanning a total range of $0.009<\lambda^2<0.055$ m$^2$. This approach yields a rotation measure spread function (RMSF; \citealp{BdB2005}) with the lowest possible sidelobes, possessing a full-width half-maximum (FWHM) of 75 rad m$^{-2}$. For each imaged channel, the dirty maps were generated using Briggs robust $=0$ weighting \citep{Briggs1995}. This weighting scheme represents a trade-off between obtaining high spatial resolution, and achieving sensitivity to large-scale structure in the lobes. The dirty images were deconvolved with the \textsc{miriad} task \textsc{pmosmem} until the RMS noise in the residual image achieved our set threshold of 1.3 times the theoretical value --- the lowest factor at which the algorithm reliably converged. The per-channel sensitivity in the resulting images is typically 0.5--1 mJy beam$^{-1}$ for Stokes $I$, 0.2--0.5 mJy beam$^{-1}$ for Stokes $Q$ and $U$, and 0.15--0.3 mJy beam$^{-1}$ for Stokes $V$, depending on frequency and location in the image. The images were then restored and convolved to a common frequency-independent resolution, re-sampled to a common pixel scale and coordinate grid, then formed into Stokes $I$, $Q$, $U$, and $V$ datacubes. The images in each frequency channel have five pixels across a $20"\times30"$ synthesized beam (in RA and Decl. respectively) --- a resolution well-matched to the scale-size of the patchy depolarization structure that we mean to investigate.

To define the lobe boundaries, and thus the region of interest for our spectropolarimetric analysis, we created a comparatively deep map of the lobes in total intensity by co-adding the re-gridded spectropolarimetric Stokes $I$ images. The result is shown in Figure \ref{fig:FornA_zoom}. The well-known features of the system are all clearly visible --- i.e. the diffuse lobe emission, the bright core / jet region, and a faint radio bridge lying largely to the south of the core that connects the two lobes (see \citealt{Ekers1983}, \citealt{Geldzahler1984}, and \citealt{Fomalont1989} for detailed descriptions of these features). 

\begin{figure}
\centering
\hspace{-1.3cm}
\includegraphics[width=0.45\textwidth]{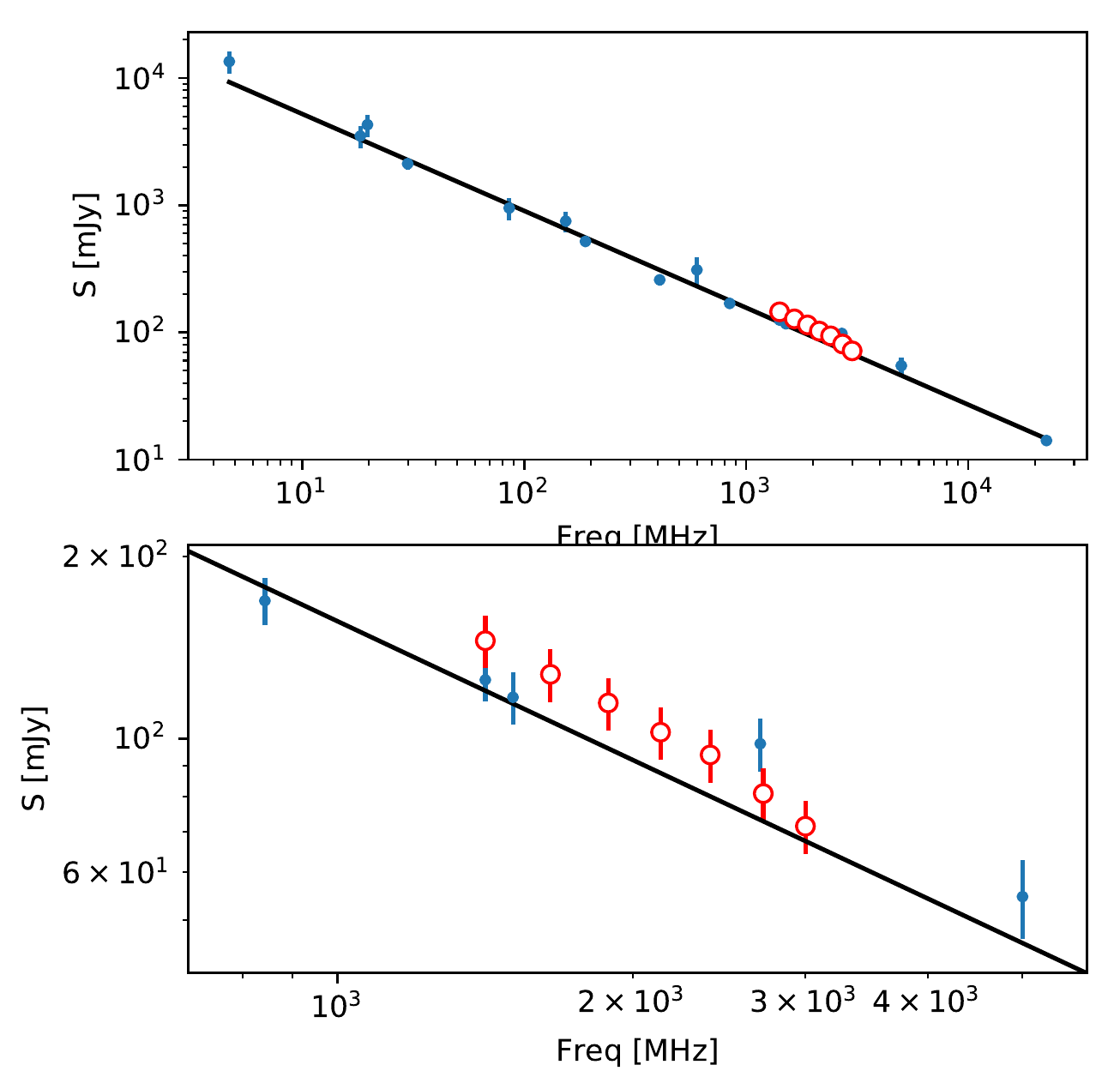}
\caption{Total flux measurements (S) of Fornax A aggregated from the literature (blue dots; \citealp{McKinley2015}), and measured from our spectropolarimetric images at selected frequencies (red open circles). The black line is a power-law fit to the \citet{McKinley2015} data only. The top panel shows the full frequency range covered by the McKinley et al. data, while the bottom panel zooms in on a more limited frequency range to show our data points more clearly. For reference, the fitted power-law model has $\alpha=-0.76$.}
\label{fig:FornAFlux}
\end{figure}

We now comment on whether our spectropolarimetric images suffer from missing flux. Figure \ref{fig:FornA_zoom} gives the impression that this might be the case, given the slight negative bowl apparent around the lobes (though its appearance is exacerbated by the arcsinh stretch employed therein). For GHz-frequency, multi-GHz-bandwidth observations of extended objects, the zero-spacing problem will affect the highest frequencies foremost, and Stokes $I$ to a much greater degree than the linearly-polarized emission, since the former is positive-definite, and will generally have a larger characteristic scale size. For our observations though, the shortest effective baseline for an object at zenith for our jointly-imaged data (see \citealp{ER1979}) is 9 m, and shorter still once baseline foreshortening is accounted for. This provides us with sensitivity to emission size scales of $\sim50$ arcminutes, which is marginally larger than the angular scale of the long axis of the entire system in Stokes $I$. However, a better estimate for the characteristic largest angular scale of the system is $\sim15$--$20$ arcminutes, since this range approximately corresponds to the angular size of each lobe, \emph{and} their separation from one another (Figure \ref{fig:FornA_zoom}). Thus, we expect the amount of missing flux to be negligible in all Stokes parameters at all frequencies. To check this, we measured the total flux in our spectropolarimetric Stokes $I$ images at several frequencies, then compared these values to measurements of total flux collated from the literature by \citet{McKinley2015}, each of which are not expected to suffer from missing flux. The results are plotted in Figure \ref{fig:FornAFlux}, along with a power-law fit to the McKinley et al. data. Our measured total fluxes are consistent with those reported in the literature --- if anything, we marginally overestimate the total flux by comparison. Moreover, we were unable to uncover evidence that this situation changed with frequency to better than 10\% of Stokes $I$, implying that our fractional Stokes $q$ and $u$ values (see Section \ref{analysis}) are also good to at least this level. Thus, we do not consider that missing flux affects our results in any important way. 


\section{Analysis}\label{analysis}

Our analysis focuses on the linearly polarized emission from Fornax A. A given state of linear polarization can be represented as a complex vector $\boldsymbol{P}$, which is related to the Stokes parameters $Q$ and $U$, the polarization angle $\psi$, the fractional polarization $p$, and the total intensity $I$, as:
 
 \begin{equation}
\boldsymbol{P} = Q + iU = pIe^{2i\psi}
\label{eq:ComplexPolVec}
 \end{equation}

\noindent Having been emitted from a point $L$, linearly polarized radiation can be Faraday rotated by magnetized plasma along the path to the observer by an amount equal to

  \begin{equation}
\Delta\psi= \phi\lambda^2
\label{eq:rotation}
 \end{equation}

\noindent where $\psi$ is the polarization angle, $\lambda$ is the observing wavelength, and $\phi$ is the Faraday depth, given by  

 \begin{equation}
\text{$\phi$($L$)} = 0.812 \int_{L}^{\text{telescope}} n_{e,\text{lobe}}\boldsymbol{B}.\text{d}\boldsymbol{s}~\text{rad m}^{-2}
\label{eq:FaradayDepth}
 \end{equation}
 
\noindent and $n_{e,\text{lobe}}$ [cm$^{-3}$] \& $\boldsymbol{B}$ [$\mu$G] are the thermal electron density and magnetic field along the line of sight (LOS) respectively, and the displacement $\boldsymbol{s}$ is in units of parsecs.

To remove changes in the linearly polarized intensity caused by simple spectral index effects, we applied our spectropolarimetric analysis to the fractional Stokes quantities $q=Q/I$ and $u=U/I$. These were calculated from the Stokes $I$, $Q$, and $U$ datacubes by:

\begin{enumerate}
\item Fitting a power law model to log($I$) vs. log($\nu$) at each pixel location, then converting this to an $I(\lambda^2)$ model.
\item Dividing the Stokes $Q(\lambda^2)$ and $U(\lambda^2)$ values by the $I(\lambda^2)$ model.
\end{enumerate}

\thispagestyle{empty}
\begin{figure*}
\centering
\includegraphics[width=0.75\textwidth]{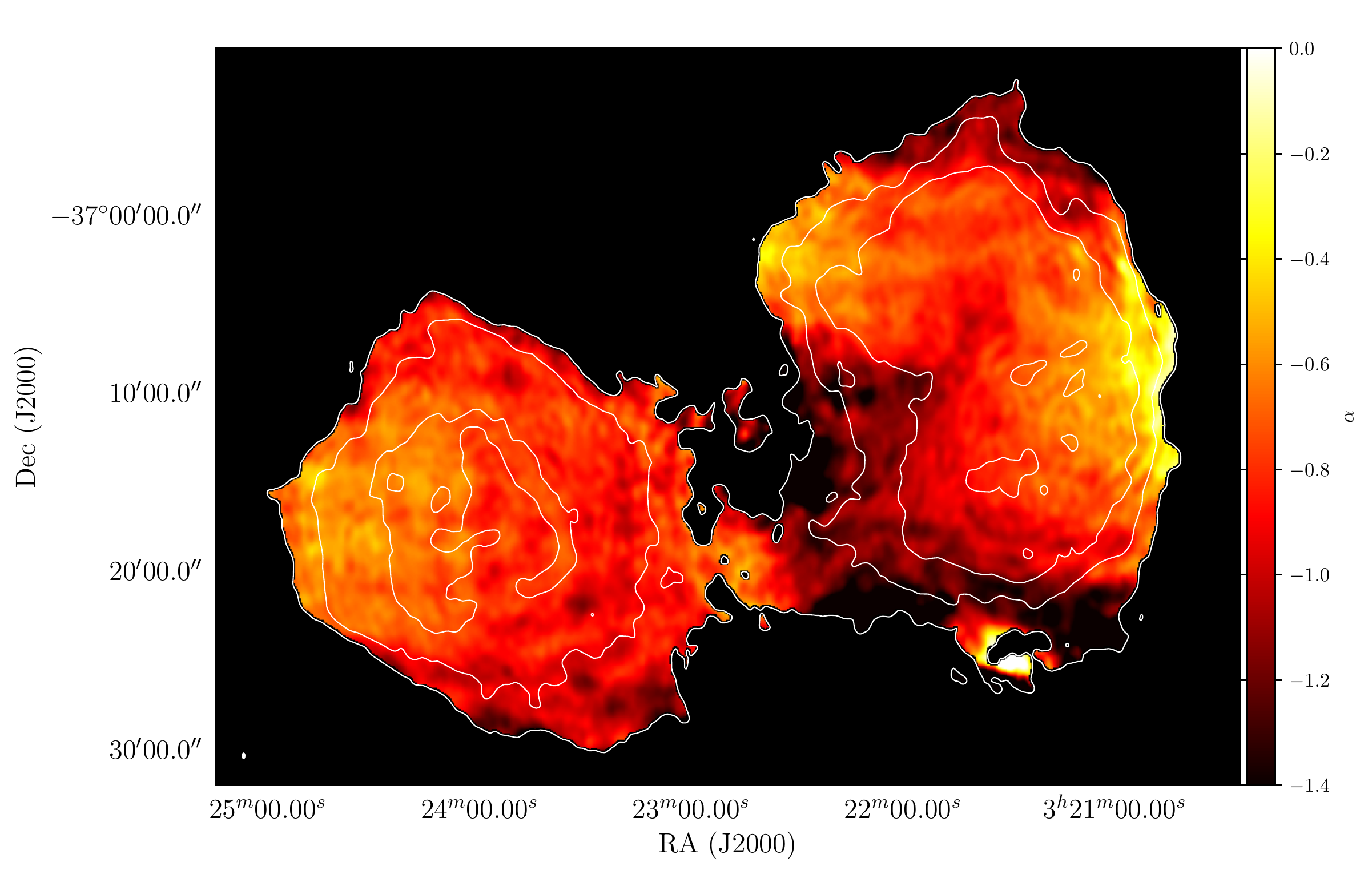}
\caption[Map of the spectral index over Fornax A ($20"\times30"$ spatial resolution)]{A colormap of the spectral index over Fornax A ($20"\times30"$ spatial resolution in RA and Decl. respectively), discussed in Section \ref{analysis}. The white contours correspond to the Stokes I intensity as shown in Figure \ref{fig:FornA_zoom}. The contour levels start at 8 mJy beam$^{-1}$, increasing by factors of 2.}
\label{fig:FornAalpha}
\end{figure*}

A map of spectral index over the lobes is presented in Figure \ref{fig:FornAalpha}, apparently revealing rich structure. We note that when convolved to a spatial resolution that matches that of the Murchison Widefield Array (MWA) --- i.e. $\sim3\times3$ arcminutes --- the small- and large-scale features apparent in the resulting map are very similar to those observed in spectral tomography maps presented by \citet{McKinley2015} (calculated between 154 MHz and 1510 MHz; cf. Figs. 5 and 6 of that work). McKinley et al. tentatively attributed this structure to imaging and deconvolution artifacts; we propose that it may in fact be real. However, since we find no clear evidence of a spatial relationship between spectral index and fractional polarization in our maps (see Figure \ref{fig:FornADepolRegions}), we defer a deeper investigation and physical analysis of the spectral index structure to paper 2; here our discussion of spectral index extends only so far as it illuminates the systematic uncertainties in our Stokes $I$ fitting, and therefore the calculation of the fractional quantities Stokes $q$ and $u$.

The spectral index is precisely determined: At each fitted pixel location for which the band-averaged Stokes $I$ intensity was greater than 16 mJy beam$^{-1}$ (i.e the vast majority of the lobes; see contours in Figure \ref{fig:FornAalpha}), the typical $1\sigma$ uncertainty in the fitted spectral index is only $\pm0.01$, while the beam-to-beam variance in spectral index is $\sigma_\alpha^2\approx5\times10^{-3}$. This relatively high level of precision is a direct result of the lever arm provided by our broad, densely-sampled 2 GHz bandwidth (see the data and fitting results plotted in Section \ref{sec:FreqDepPol}). We conclude that uncertainties in the Stokes $I$ spectral model have negligible impact on our fitting results and analysis. Moreover, a power-law fit provides an excellent description of the Stokes $I$ spectral behavior (again, see plots presented in Section \ref{sec:FreqDepPol}). Nevertheless, we tested the the effect of deriving Stokes $q$ and $u$ by fitting a 2nd degree polynomial to the log($I$) vs. log($\nu$), and by dividing Stokes $Q$ and $U$ by Stokes $I$ on a point-by-point basis. In both cases, the effect on the Stokes $q$ and $u$ data and subsequent results was also negligible.

We applied two complementary spectropolarimetric analysis techniques to the Stokes $q$ and $u$ data. The first is the combined RM synthesis \citep{BdB2005} + \textsc{rmclean} \citep{Heald2009} technique, which generates the so-called Faraday dispersion spectrum (FDS) --- a direct reconstruction of the so-called Faraday dispersion function (FDF) from the observational data, itself a complex vector that specifies the intensity and intrinsic emission angle of linearly-polarized emission as a function of Faraday depth along the line of sight. We applied it to the Stokes $q$ and $u$ datacubes at all pixel locations where the Stokes $I$ flux density exceeded 8 mJy beam$^{-1}$ (see Figure \ref{fig:FornA_zoom}). This cutoff level was selected to include the main morphological features of the system, whilst excluding peripheral regions with a low polarized signal-to-noise ratio (S/N). After calculating the FDS at each pixel location, we recorded:

\begin{enumerate}
\item $p_{\text{FDS}}$: The maximum amplitude of $|$FDS$|$. This provides a measure of the band-averaged fractional polarization (plus Ricean bias, which is insignificant here --- see \citealp{Hales2012}), assuming that the FDS is dominated by emission from a single Faraday depth. As we show in Section \ref{sec:FreqDepPol}, this assumption is often not justified for this object, and in such cases, $p_{\text{FDS}}$ provides a lower limit on the band-averaged fractional polarization along a given sight-line. 
\item $\phi_{\text{peak}}$: The peak Faraday depth --- i.e. the Faraday depth at which $|$FDS$|$ is maximized. This represents the Faraday depth from which the bulk of linearly-polarized emission has emerged from along a particular sight line. 
\item The noise level in the FDS
\end{enumerate}
 
RM synthesis + \textsc{rmclean} provides an excellent estimate of both the band-averaged fractional polarization, (with a typical beam-to-beam variance of $\sigma_p^2\approx4\times10^{-4}$ for our data), and the peak Faraday depth, with a theoretical uncertainty of $\lesssim2$ rad m$^{-2}$ for an RMSF width of 75 rad m$^{-2}$ and a band-averaged polarized signal-to-noise ratio (S/N) $\geq8$ --- see Figure \ref{fig:FornAFDErrRegions}), as well as the range of Faraday depths over which a radio source emits. Moreover, the FDS is calculated without supplying \emph{a priori} constraints on the nature of the Faraday rotation with physical models, and thereby acts as a point of comparison for the best-fit Faraday rotation models obtained through our second primary analysis technique --- ($q$,$u$)-fitting. 

($q$,$u$)-fitting provides a complementary, and in some respects superior \citep{Sun2015}, diagnostic of the nature and structure of magnetized plasmas. This technique has the advantage that physically-motivated models can be fit directly to the polarization data, directly yielding physical constraints on the magneto-ionized structure of sources where a unique best-fit model can be shown to exist. Moreover, ($q$,$u$)-fitting avoids subtle issues associated with the output of RM synthesis and {\sc rmclean} (e.g. the inability to localize individual {\sc rmclean} components within the width of the RMSF for sources that are resolved in Faraday depth space). 

We fit all combinations of models consisting of one or two polarized emission components, with each described by a variant (described below) of the following function (\citealp{OSullivan2017}, see also \citealp{Burn1966,Sokoloff1998}):  

\begin{eqnarray}
\boldsymbol{P}_j(\lambda^2) = p_{0[j]}e^{2i({\psi_{0[j]}}+\text{RM}_{[j]}\lambda^2)}e^{-2\sigma_{\text{RM}[j]}^2\lambda^4}\text{sinc}(\Delta \phi_{[j]}\lambda^2)
\label{eqn:ShaneMod}
\end{eqnarray}

\noindent where $p_{0[j]}$, $\psi_{0[j]}$, and RM$_{[j]}$ are the initial fractional polarization, initial polarization angle, and rotation measure of the $j$th emission component (respectively), and $\sigma_{\text{RM}[j]}$ and $\Delta \phi_{[j]}$ characterize Faraday-dispersive effects. The first exponential term in this equation models pure Faraday rotation, the second exponential term models depolarization by a turbulent magnetoionic foreground (i.e. `external Faraday dispersion'; but see \citealt{Burn1966} and \citealt{Sokoloff1998} for discussion of different scenarios in which the depolarization behavior can sometimes differ only subtly from the effect modeled here), while the last term is capable of modeling depolarization behavior from mixed synchrotron-emitting and Faraday-rotating plasma (which we refer to as internal differential Faraday rotation), among other possible scenarios \citep{Schnitzeler2015}. 

The different variants of Eqn. \ref{eqn:ShaneMod} just alluded to alternately remove the external Faraday dispersion term, the internal differential Faraday rotation term, and both depolarization terms, resulting in emission components described by Faraday rotation + internal differential Faraday rotation, Faraday rotation + external Faraday dispersion, and Faraday rotation-only respectively. We also fit a three emission component model with all of the terms in Eqn. \ref{eqn:ShaneMod} included, in order to adequately capture the behavior of particularly complex emission. 

We fit each such model (which we now refer to as `model types') to the data using the procedure detailed by \citet{Anderson2016}. To summarize, we used the {\sc emcee} sampler \citep{FM2013} to identify parameter values that maximize the following likelihood function for polarization data ($q_i$,$u_i$) and a model (q$_{mod,i}$,u$_{mod,i}$) (where $i$ is the channel index):

\begin{equation}
\mathcal{L} = \prod\limits_{i=1}^n \frac{1}{\pi\sigma_{q_{i}}\sigma_{u_{i}}}\text{exp}\Bigg(-\frac{(q_i-q_{mod,i})^2}{2\sigma_{q_{i}}^2}-\frac{(u_i-u_{mod,i})^2}{2\sigma_{u_{i}}^2}\Bigg)
\label{eq:likelihood}
\end{equation}

We assigned uniform prior PDFs to all parameters over physically reasonable ranges --- i.e. [$-\pi/2$,$\pi/2$) rad for $\psi_0$, [0,0.8] for $p_0$, [-3000,3000] rad m$^{-2}$ for the RM, and [0,1500] rad m$^{-2}$ for each of $\sigma_{\text{RM}}$ and $\Delta \phi$ --- and set the PDF to zero over the remaining parameter space. 

In order to improve the computational tractability of the ($q$,$u$)-fitting analysis, we down-sampled the Stokes $q$ and $u$ datacubes to one pixel per synthesized beam area. We then fit each of the model types described above to the ($q$,$u$) data at pixel locations where $p_{\text{FDS}}<0.2$ in the down-sampled cube --- i.e. inside the low-$p$ patches --- and at select pixel locations satisfying $0.4\leq p_{\text{FDS}}\leq0.65$ to provide a control sample. 

We determined the relative merit of the best-fitting model of each type at a given pixel location using the Akaike information criterion (AIC; \citealp{Akaike1974}), calculated as: 

\begin{equation}
\text{AIC}_M = 2k-2\text{ln}(\mathcal{L}_{\text{max}})
\label{eq:AIC}
\end{equation}

Here, $k$ is the number of fitted model parameters, and $\mathcal{L}_{\text{max}}$ is the maximum of the likelihood distribution for a model $M$. For two models $M1$ and $M2$, $M2$ is exp((AIC$_{M1}-$AIC$_{M2}$)/2) times as likely as $M1$ to minimize the information lost by under- or over-fitting the data. We considered a model $M1$ to be strongly favored over a model $M2$ when AIC$_{M1}+ 10 <$ AIC$_{M2}$ (i.e. $>99$\% confidence). 

Finally, we note that an explicit and detailed comparison of the output of RM synthesis and ($q$,$u$)-fitting is beyond the scope of this paper --- we will present such analysis in paper 2.

\section{Results}\label{results}

\subsection{The low-$p$ patches: Basic morphology and attributes}\label{sec:overviewpol}

In Figure \ref{fig:FornADepolRegions}, we present a map of $p_{FDS}$ over the lobes. In this and subsequent figures, we use polygons labeled A through R to delineate individual complexes of low-$p$ patches, which we refer to as `boxes'. The low-$p$ patches stand out as the dark regions, though we note that the band-averaged linearly-polarized S/N remains $\geqslant8$ inside the outer lobe boundary. The appearance of these patches in Figure \ref{fig:FornADepolRegions} (i.e. in broadband fractional polarization) is much the same as in Figure 1 of \citet{Fomalont1989} (i.e. a narrowband VLA map in linearly-polarized intensity at 1.5 GHz and 14" resolution), and thus, despite the differences in our respective observational setup and capabilities, we are analysing the same structures described by those authors. 

The low-$p$ patches are resolved in the transverse direction along most of their length, typically by more than three synthesized beam-widths at $20"\times30"$ spatial resolution. This may not be immediately obvious in Figure \ref{fig:FornADepolRegions}, so in Appendix A (Figures \ref{fig:FornAFDZoomRegionsDP_paper} and \ref{fig:FornAFDZoomRegionsFD_paper}), we present RM synthesis/\textsc{rmclean}-derived maps of $p_{\text{FDS}}$ and $\phi_{\text{peak}}$ over selected low-$p$ patches at $4"\times8"$ resolution. The $uv$ tapering scheme required to attain this spatial resolution reduces the polarized S/N markedly, precluding the use of these higher resolution images as the main basis of our analysis. Nevertheless, the low-$p$ patches are clearly resolved in these maps, and their widths, morphologies, level of fractional polarization, and indeed Faraday depth structure (see Section \ref{sec:spatialFD}), do not appear to change appreciably despite the substantially increased spatial resolution. Thus, the low-$p$ patches are real structures with a physical extent that approximately corresponds to their angular scale in Figure \ref{fig:FornADepolRegions}.

In contrast to the main body of the lobes, where $p_{FDS}$ (expressed as a percentage) is typically $\sim30\%$ and in some places higher than $\sim60\%$, the value of $p_{FDS}$ within the low-$p$ patches is typically less than 10\%, and reaches as low as 0.8\%. Thus, if the low-$p$ patches are associated with a Faraday depolarising medium, it must extend close to the lobe surface --- i.e. to within a small fraction of one lobe diameter. Were this not the case, synchrotron emission from intervening lobe material would most likely boost the level of observed linear polarization to values more in line with those seen elsewhere in the lobes. 

Finally, we note that while the low-$p$ patches are not generally associated with known foreground objects \citep{Fomalont1989}, there are two exceptions in the form of small depolarization silhouettes. The first is generated by the Sc galaxy NGC 1310 (\citealp{Fomalont1989,SF1992}), and is located in box Q. The second --- dubbed `the ant' --- is associated with an extragalactic cloud of ionized hydrogen \citep{BH1995}. While its morphology is poorly resolved at $20"\times30"$ spatial resolution in band-averaged fractional polarization, it can nevertheless be discerned at the north-west corner of box K. An updated analysis of these structures will be presented elsewhere.

\subsection{Spatial-polarimetric analysis of the low-$p$ patches}\label{sec:spatiopol}

\subsubsection{Associated structure in peak Faraday depth}\label{sec:spatialFD}

In Figure \ref{fig:FornAFDRegions}, we present a map of $\phi_{\text{peak}}$ over the lobes. The structure present therein is both obvious and remarkable. It typically manifests as curved, elongated interfaces across which the peak Faraday depth varies rapidly, often changing sign in the process (see boxes C, K, and P for clear examples of this morphology). In some cases though, the interface is not elongated but point-like (i.e. the interface between positive and negative Faraday depth is around one synthesized beamwidth in linear extent --- see isolated features in boxes D, E, and L for examples).  

\begin{figure*}
\centering
\includegraphics[width=0.75\textwidth]{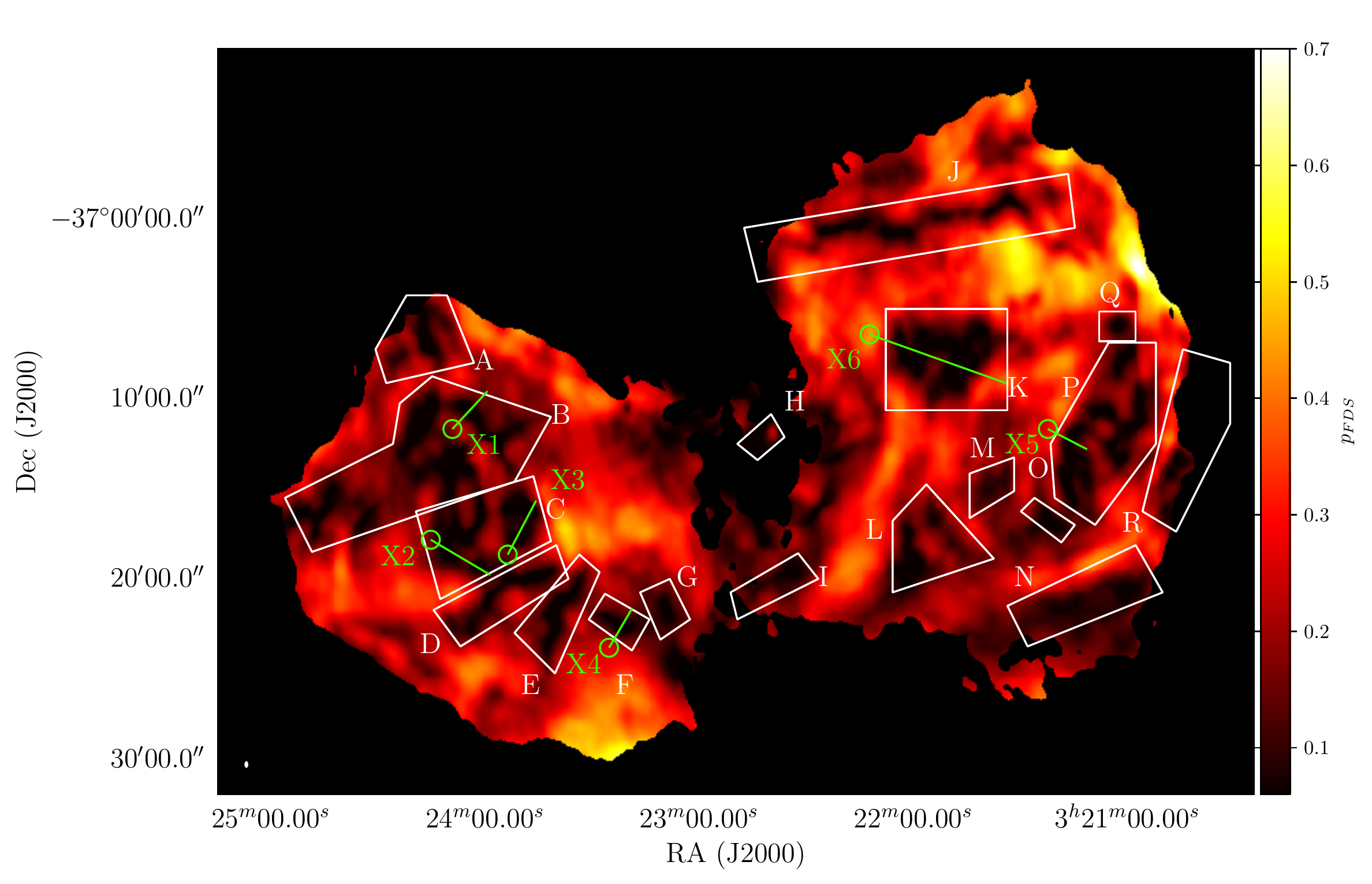}
\caption[Map of fractional polarization over Fornax A ($20"\times30"$ spatial resolution)]{A map of the magnitude of the RM synthesis $+$ \textsc{rmclean}-derived FDS (i.e. band-averaged fractional polarization) of Fornax A ($20"\times30"$ spatial resolution in RA and Decl. respectively). Depolarized patches are seen as darker areas within the radio lobes. Note that the lower image scale cutoff has been set to 0.05 to more clearly show these features. We have indicated the most heavily depolarized complexes with boxes labeled A--R for ease of referencing. Note that region Q is a depolarization silhouette generated by the foreground Sc galaxy NGC 1310 \citep{Fomalont1989,SF1992}. The synthesized beam size is indicated in white in the bottom left corner. The green lines  indicate the cross-sections plotted in Figure \ref{fig:filamentaryMyDearWatson}, with the attached open circles indicating the start of the cross-section (see Section \ref{sec:XSections}).}
\label{fig:FornADepolRegions}
\end{figure*}

\begin{figure*}
\centering
\includegraphics[width=0.75\textwidth]{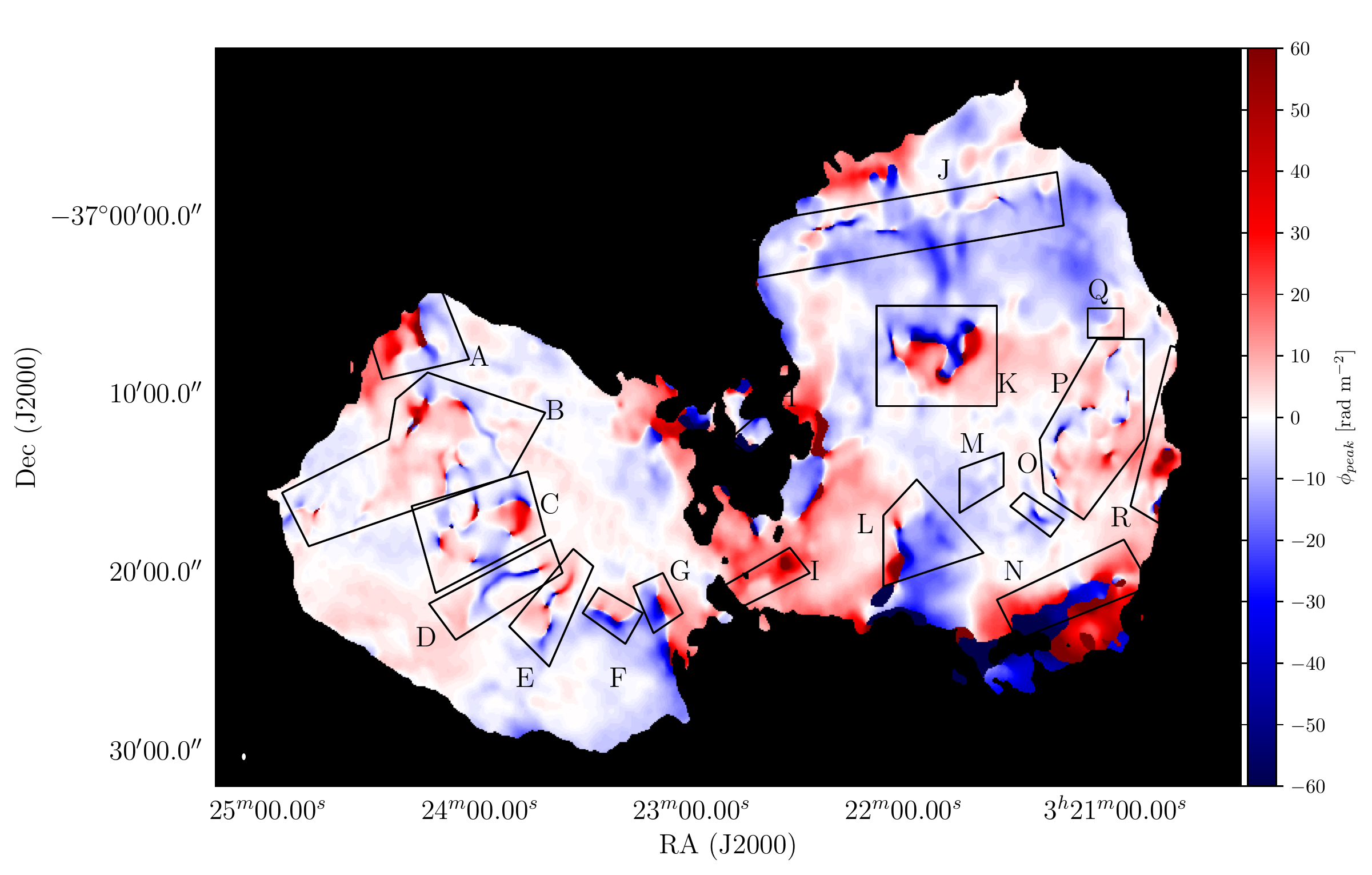}
\caption[Map of peak Faraday depth over Fornax A ($20"\times30"$ spatial resolution)]{A map of the peak Faraday depth of Fornax A, measured over 1.28--3.1 GHz with a spatial resolution of $20"\times30"$, corrected for a uniform Galactic RM foreground contribution of $-6$ rad m$^{-2}$ (see Section \ref{sec:spatialFD}). Note that the color map saturates at $\phi_{\text{peak}}=\pm60$ rad m$^{-2}$. The boxes A--R enclose the same regions as those in Figure \ref{fig:FornADepolRegions}. The white boxed ellipse in the lower left indicates the synthesized beam size.}
\label{fig:FornAFDRegions}
\end{figure*}

From a visual comparison of the $p_{\text{FDS}}$ and $\phi_{\text{peak}}$ maps (Figs. \ref{fig:FornADepolRegions} and \ref{fig:FornAFDRegions} respectively), it is clear that a) all of the interfaces in Faraday depth lie within low-$p$ patches, generally near their central region or `spine', and that b) the low-$p$ patches extend substantially beyond the immediate vicinity of the interfaces alone, and appear to occur wherever either the magnitude of $\phi_{\text{peak}}$, or spatial gradients in its value, are enhanced relative to the positions in the lobe outside the low-$p$ patches. Thus, the patches in question are spatially coincident with distinctive structure in peak Faraday depth, which strongly suggests that they are generated by Faraday effects along the line of sight to the lobes.

The magnitude of the Faraday depth is considerably higher near the interfaces than elsewhere in the lobe --- the mean for $p_{\text{FDS}}<0.1$ is $\sim25$ rad m$^{-2}$, and values up to $\sim$60 rad m$^{-2}$ are common. Note that the uncertainty on $|\phi_{\text{peak}}|$ ranges from $\sim2$--6 rad m$^{-2}$ in regions with moderate to low fractional polarization (respectively; Figure \ref{fig:FornAFDErrRegions}), so that the extreme RM values previously reported towards the system of $-15$ and $+10$ rad m$^{-2}$ by \citet{Fomalont1989} are significantly exceeded, mostly in the immediate vicinity of the interfaces. This discrepancy is probably due to the fact that $p(\lambda^2)$ exhibits considerable frequency-dependent structure at these locations, often including strong depolarization (see Section \ref{sec:FreqDepPol}). Given that the Galactic foreground RM at the location of Fornax A is only $\sim-6$ rad m$^{-2}$ (Eqn. 12 of \citealt{Anderson2015}), the sense of the dominant, regular, line-of-sight component of the magnetic field must change across the interfaces.

\thispagestyle{empty}

\thispagestyle{empty}

\subsubsection{Associated structure in the projected magnetic field orientation}\label{sec:spatialB}
 
The observed relationship between fractional polarization and peak Faraday depth motivated us to consider how these quantities (i.e. $p_{\text{FDS}}$ and $\phi_{\text{peak}}$), as well as morphological features in the Stokes $I$ emission, are spatially related to the sky-projected, emission-weighted magnetic field orientation in the lobes. We denote this latter quantity $\vartheta_{\boldsymbol{B}}$, which we define as:

\begin{equation}
\vartheta_{\boldsymbol{B}}=\frac{1}{2}\times\text{tan}^{-1}\bigg( \frac{u(0)_{\text{bf}}}{q(0)_{\text{bf}}} \bigg) + \pi/2
\label{eq:initpolangle}
\end{equation}

\noindent where $q(0)_{\text{bf}}$ and $u(0)_{\text{bf}}$ are the values of $q(\lambda^2)$ and $u(\lambda^2)$ predicted by the best-fit polarization model for a given pixel location evaluated at $\lambda^2=0$ m$^2$. Before continuing, we caution that extrapolating these models from $\lambda^2=0.01$ rad m$^{-2}$ may be unreliable in regions where the frequency-dependent polarization behavior is complex, as it can be near the low-$p$ patches (see Section \ref{sec:FreqDepPol}). It is difficult to predict and quantify these effects; future observations at higher frequencies will be required to constrain the intrinsic polarization angle with greater accuracy in such regions. 

\begin{figure*}
\centering
\hspace{-3.3cm}
\includegraphics[width=0.8\textwidth]{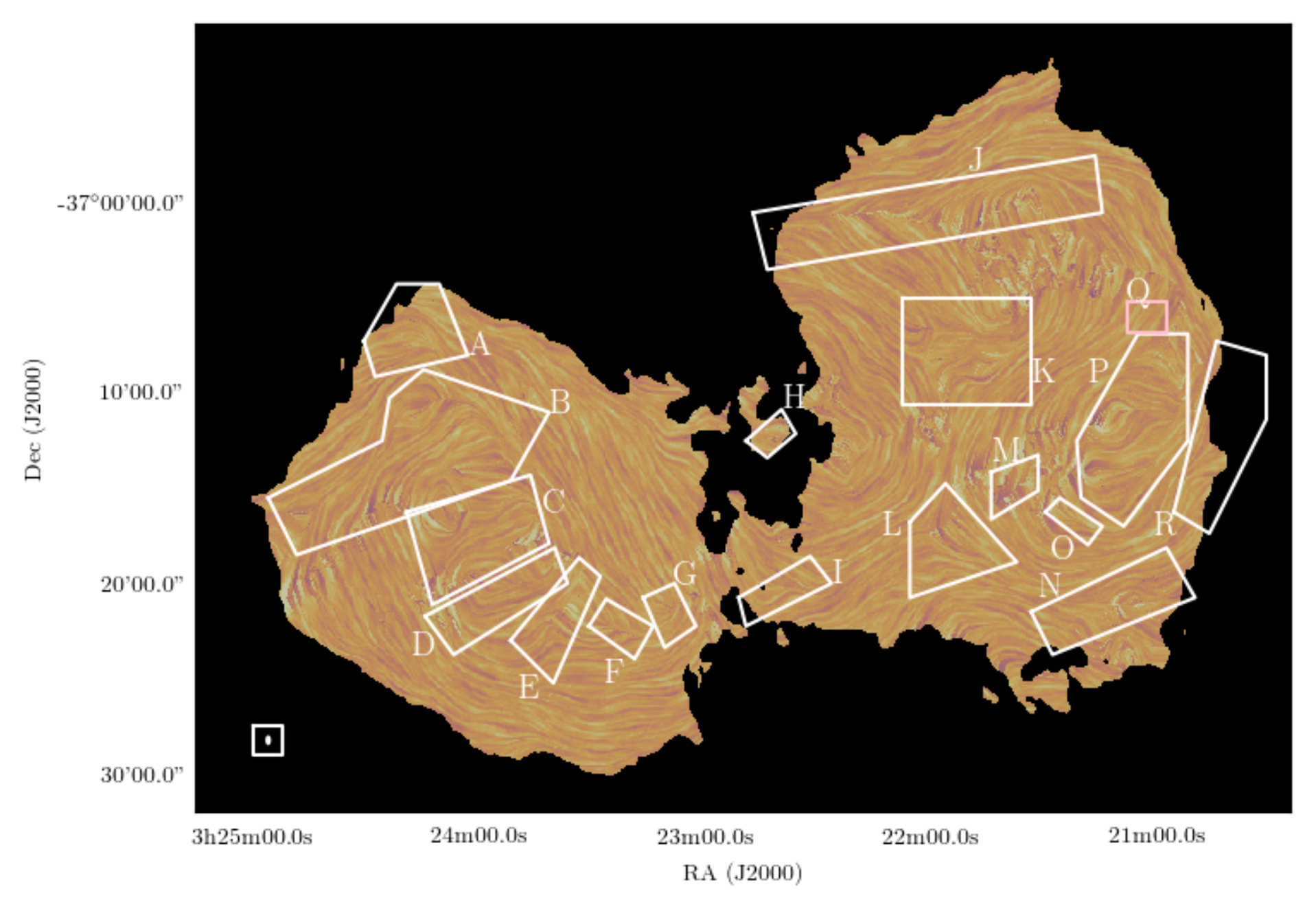}
\caption[]{Line integral convolution (LIC) map of the sky-projected magnetic field orientation, $\vartheta_{\boldsymbol{B}}$. The textured lines run parallel to $\vartheta_{\boldsymbol{B}}$ at each position in the map, by construction (see main text). The noisy region above boxes K and Q results from the best fit model taking Stokes q values close to zero at $\lambda^2=0$ m$^2$. The arctan function amplifies the effects of the fitting uncertainty in these locations. The boxes are the same as those in Figure \ref{fig:FornADepolRegions}. We note that the color scale in this figure is used as a texture to trace the LIC streamlines only --- the color value at a given pixel location does not reflect any physical property of the underlying magnetic field.}
\label{fig:FornA_B_flow}
\end{figure*}

\begin{figure*}
\centering
\hspace{-1.3cm}
\includegraphics[width=0.9\textwidth]{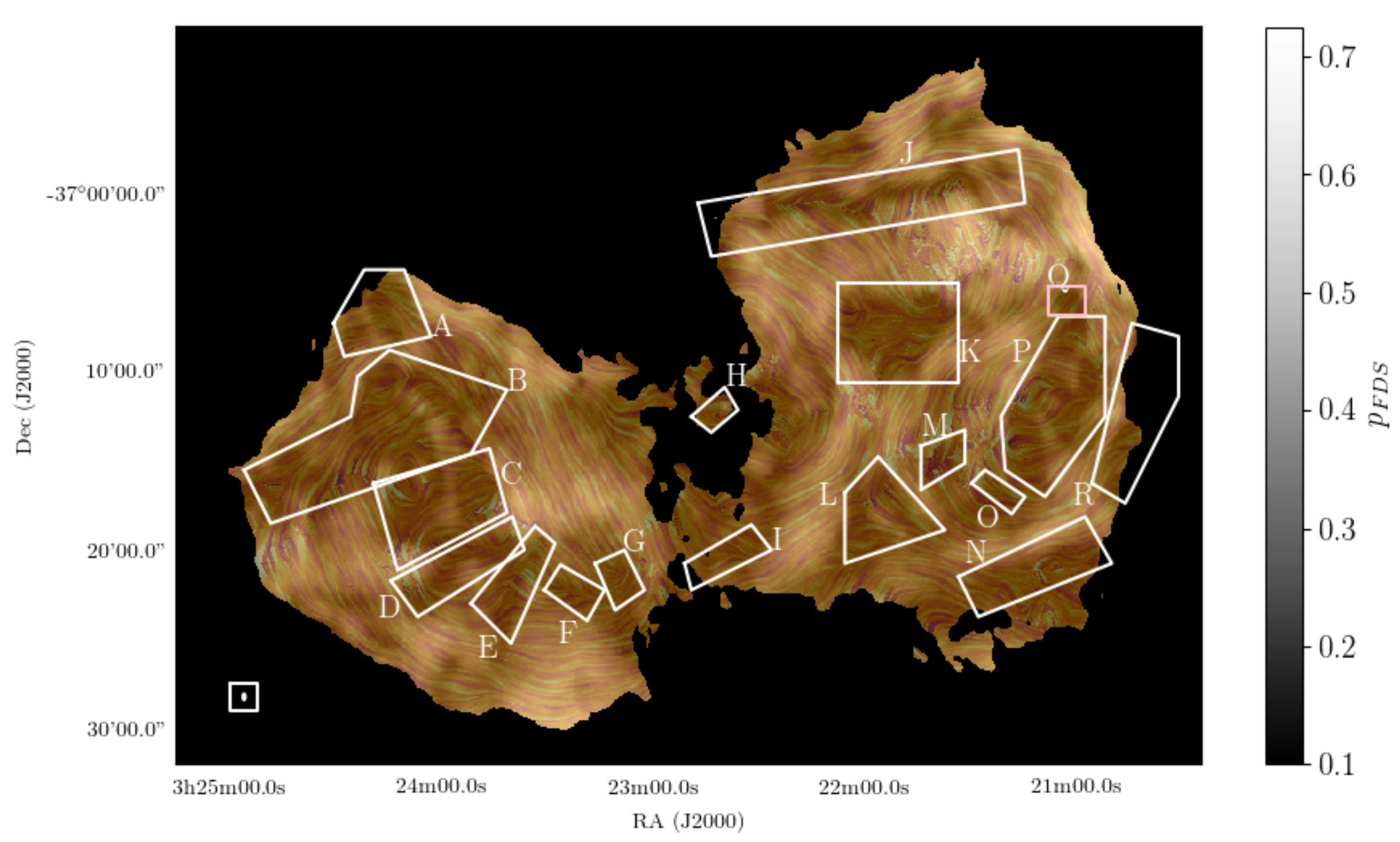}
\caption[]{As for Figure \ref{fig:FornA_B_flow}, but with LIC streamlines overlaid on a grayscale map of $p_{\text{FDS}}$.}
\label{fig:FornA_B_flow_olay_fp}
\end{figure*}

\begin{figure*}
\centering
\hspace{-1.3cm}
\includegraphics[width=0.9\textwidth]{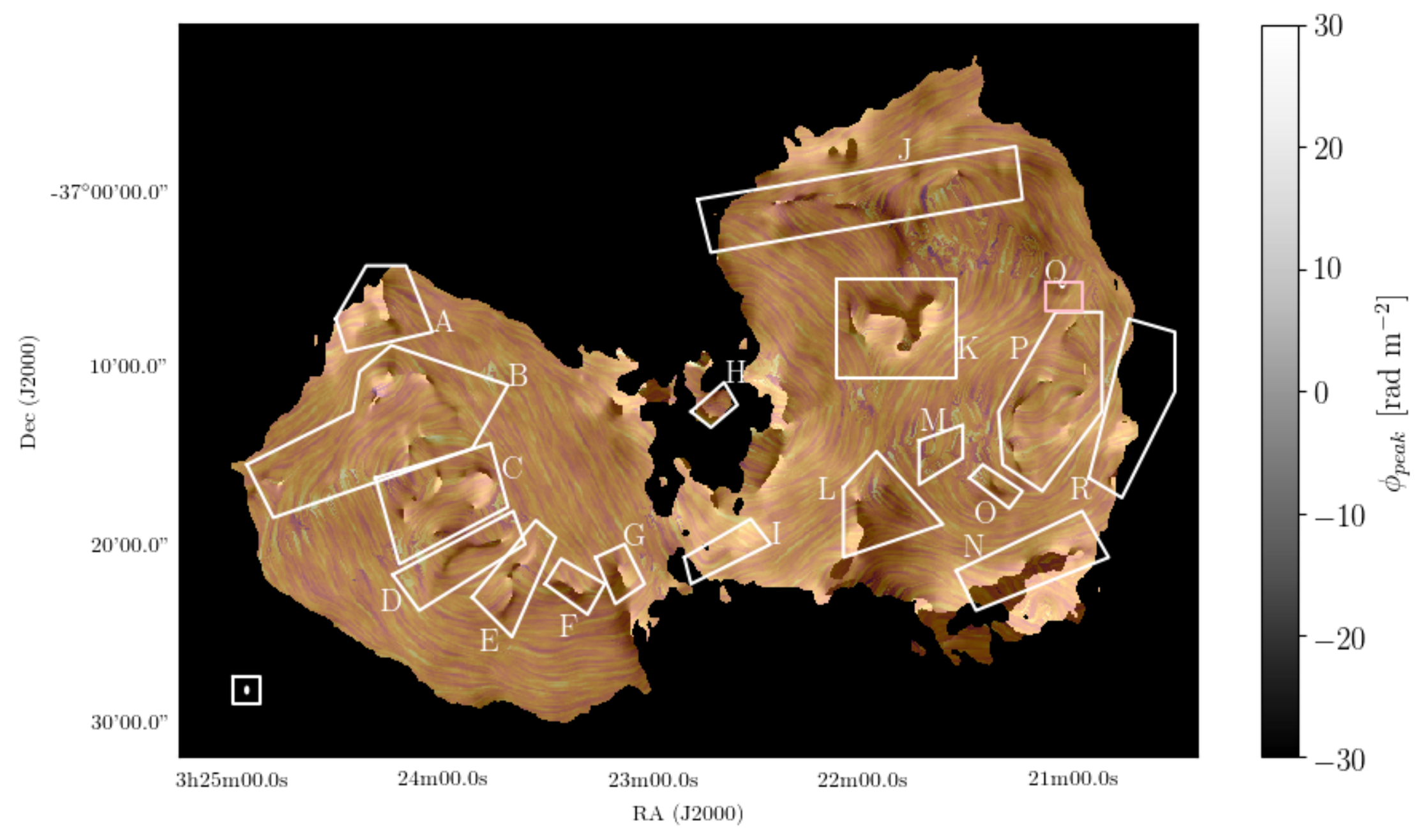}
\caption[]{As for Figure \ref{fig:FornA_B_flow}, but with LIC streamlines overlaid on a grayscale map of $\phi_{\text{peak}}$.}
\label{fig:FornA_B_flow_olay_fd}
\end{figure*}

\begin{figure*}
\centering
\hspace{-1.3cm} 
\includegraphics[width=0.9\textwidth]{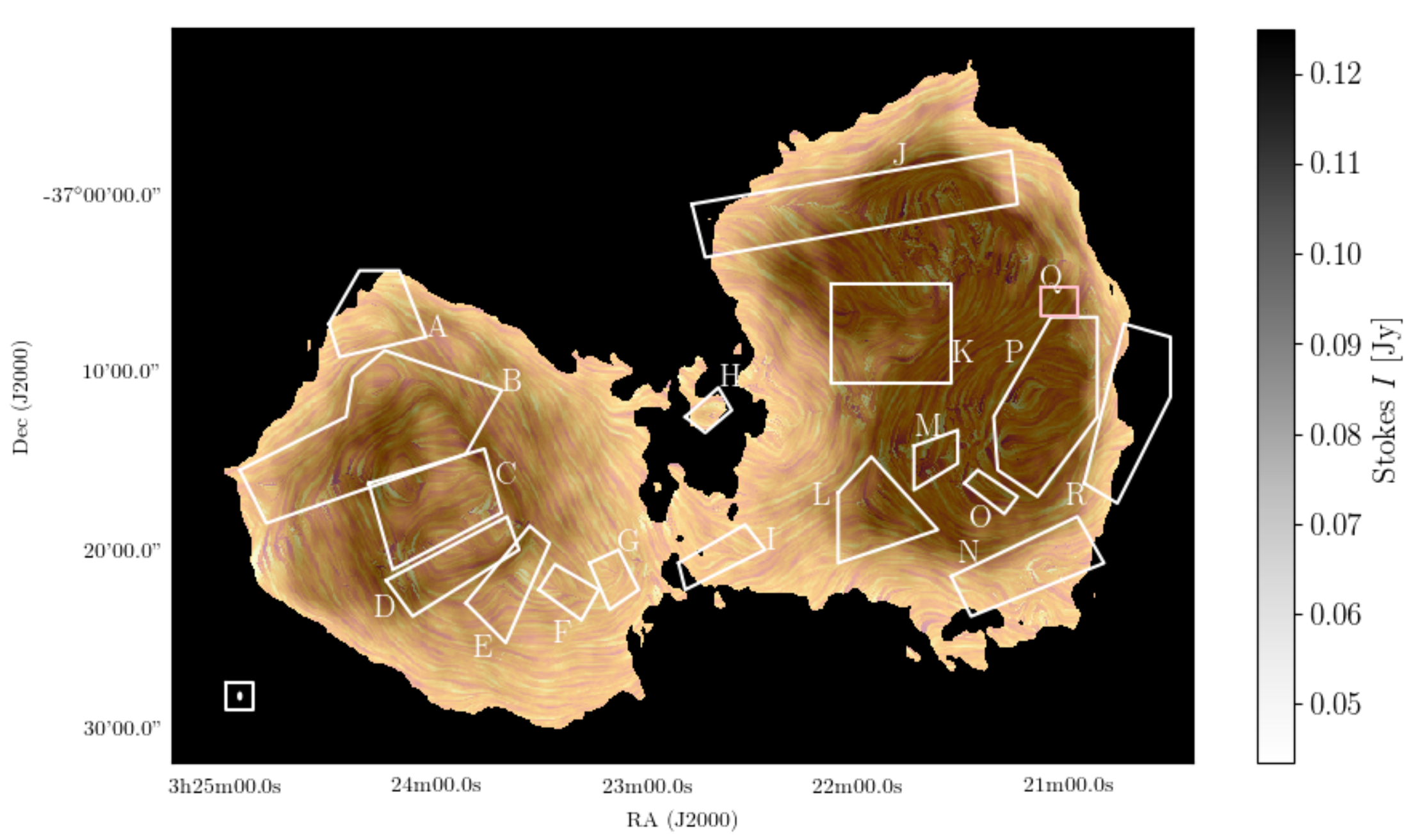}
\caption[]{As for Figure \ref{fig:FornA_B_flow}, but with LIC streamlines overlaid on a grayscale map of Stokes $I$.}
\label{fig:FornA_B_flow_olay_ci}
\end{figure*}

With this caveat, we generated a visualization of $\vartheta_{\boldsymbol{B}}$ using the line integral convolution method (LIC; \citealt{Cabral1993}), whereby a dense collection of streamlines are generated that locally follow $\vartheta_{\boldsymbol{B}}$, by construction. We present a map of the LIC streamlines in Figure \ref{fig:FornA_B_flow}. We note that the large-scale structure apparent therein matches that derived by \citet{GW1971} using data from the Parkes telescope at 6 cm. In Figs. \ref{fig:FornA_B_flow_olay_fp},  \ref{fig:FornA_B_flow_olay_fd}, and \ref{fig:FornA_B_flow_olay_ci}, we superpose the LIC streamlines on maps of $p_{\text{FDS}}$, $\phi_{\text{peak}}$, and Stokes $I$, respectively. 

Figure \ref{fig:FornA_B_flow} shows that $\vartheta_{\boldsymbol{B}}$ exhibits two distinctly different behaviors in the lobes. Some areas possess relatively smooth and coherent structure in $\vartheta_{\boldsymbol{B}}$ on length scales of $\sim50$ kpc ($\sim10$ arcminutes) and above, while in others, $\vartheta_{\boldsymbol{B}}$ is knotted into small eddy-like structures or cells on scales of $\sim25$ kpc ($\sim5$ arcminutes) and below. These structures are arranged differently in the two lobes: In the eastern lobe, the cells are largely confined to a roughly north-south oriented band (boxes A--F) wherein $\vartheta_{\boldsymbol{B}}$ is predominantly oriented east-west, while elsewhere, $\vartheta_{\boldsymbol{B}}$ predominantly aligns with the projected edge of the lobe. The structure of the western lobes is more complicated: $\vartheta_{\boldsymbol{B}}$ takes on a `figure of 8' shape on the largest scales (with part of the lower loop of the `8' truncated), while prominent small-scale cells in $\vartheta_{\boldsymbol{B}}$ are located above, below, and to the side of the crossing point in the `8' (see boxes K, L, and P respectively). Regardless, it is clear that the low-$p$ patches are primarily found in the locations where $\vartheta_{\boldsymbol{B}}$ exhibits this small-scale eddy-like structure (see Figure \ref{fig:FornA_B_flow_olay_fp}). More specifically, the low-$p$ patches are located where the orientation of $\vartheta_{\boldsymbol{B}}$ changes rapidly, and perhaps more intriguingly, in regions that are immediately adjacent to such (and also immediately adjacent bright Stokes $I$ filaments; see depolarized patches in boxes B, C for example). 

The interfaces in $\phi_{\text{peak}}$ also occur in and around the complicated intersections of streamlines associated with the $\vartheta_{\boldsymbol{B}}$ cells. More precisely, the $\phi_{\text{peak}}$ interfaces often occur where streamlines associated with the $\vartheta_{\boldsymbol{B}}$ meet at oblique angles. It might then be thought that the interfaces are merely artefacts caused by observing emission from crossed magnetic fields in the same synthesized beam. There are several arguments against this however:

\begin{enumerate}
\item The interfaces connect up smoothly to broader structures in the FD map of like sign, which extend over regions much larger than a synthesized beam area (see Figs. \ref{fig:FornAFDRegions} and \ref{fig:FornAFDZoomRegionsFD_paper}). 
\item The  $\phi_{\text{peak}}$ and $p_{\text{FDS}}$ morphological structure does not appear to change significantly with a factor of $\sim4$ improvement in spatial resolution in both RA and Decl. (see Appendix A, Figures \ref{fig:FornAFDZoomRegionsDP_paper} and \ref{fig:FornAFDZoomRegionsFD_paper}).
\item Crossed magnetic fields should generate frequency-independent depolarization, whereas we show in Section \ref{sec:FreqDepPol} that this is not generally what is observed.
\end{enumerate}

\noindent Rather, the observed $\phi_{\text{peak}}$--$\vartheta_{\boldsymbol{B}}$ spatial relationship might represent evidence that the Faraday depth interfaces occur at junctures in the bulk magnetic field structure --- and perhaps more speculatively, in bulk flows of material --- in the lobes. \\ 

In both lobes, emission features in total intensity also coincide with structure in $\vartheta_{\boldsymbol{B}}$ to some extent (see Figure \ref{fig:FornA_B_flow_olay_ci}), though this is most apparent for the large-scale component of the structure apparent in $\vartheta_{\boldsymbol{B}}$. Generally, the relationship manifests as an alignment between $\vartheta_{\boldsymbol{B}}$ and the projected edge of the lobes (as is commonly observed; e.g. \citealp{Laing1980,HE2011}), and between $\vartheta_{\boldsymbol{B}}$ and the brighter Stokes $I$ filaments. The smaller-scale cellular structure in $\vartheta_{\boldsymbol{B}}$ is generally found in between the bright Stokes $I$ filaments, and often appears to merge at oblique angles with $\vartheta_{\boldsymbol{B}}$ in the Stokes $I$ filaments (see the western regions of boxes B and C for example). It is not clear whether this is a projection effect, or whether different Stokes $I$ filaments are interconnected by the magnetic field in this way. Nevertheless, the low-$p$ patches appear to be found in-between, and adjacent to, the bright Stokes $I$ filaments. Though the precise nature of this relationship is not clear, it nevertheless implies that the process(es) that generate the depolarized patches are in some way linked to the process(es) that generate the bright Stokes $I$ filaments.

\thispagestyle{empty}

\thispagestyle{empty}

\thispagestyle{empty}

\thispagestyle{empty}

\subsubsection{Cross-sectional analysis}\label{sec:XSections}

\begin{figure*}
\centering
\includegraphics[width=1\textwidth]{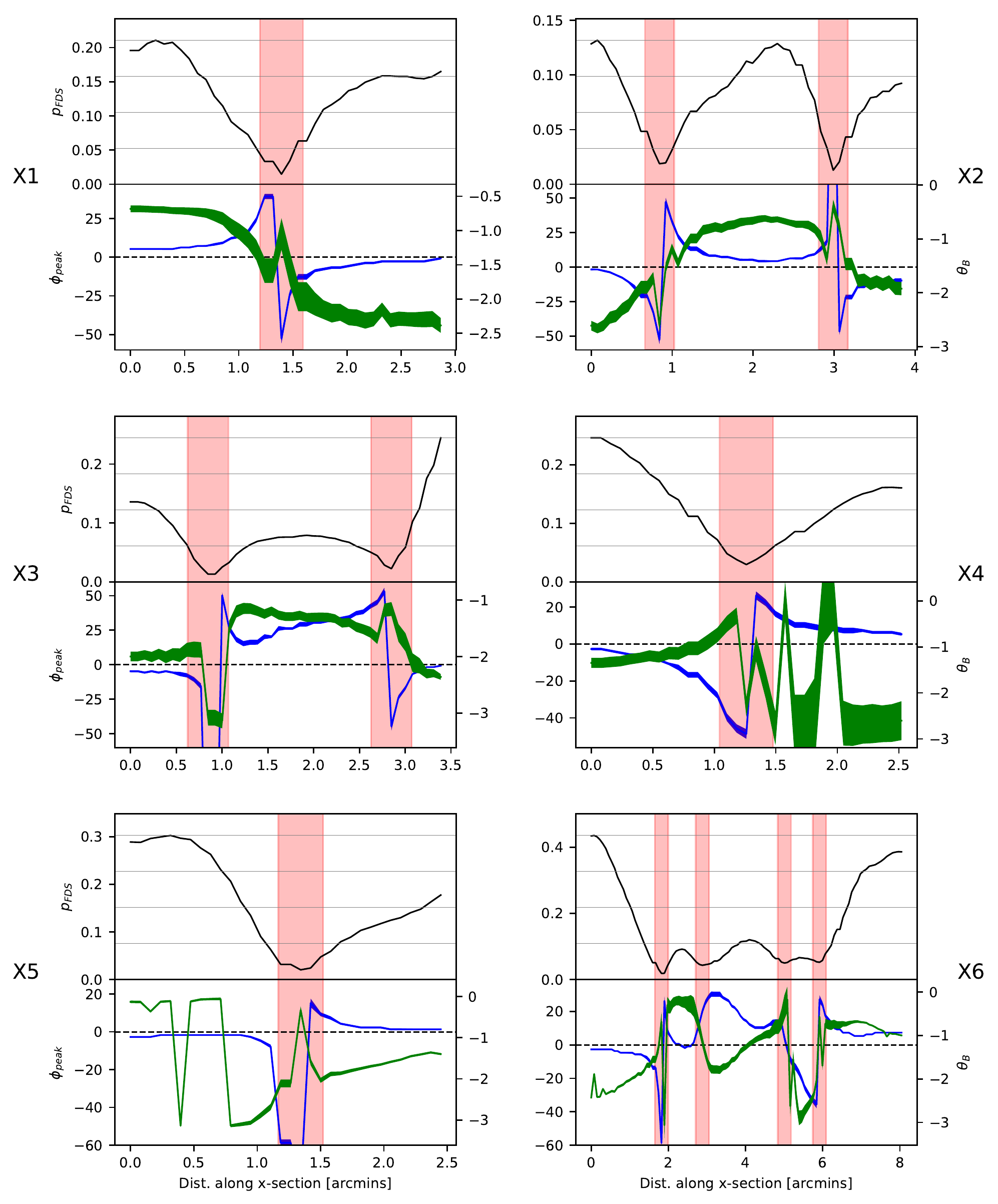}
\caption[]{Plots of $p_{\text{FDS}}$ (top panel in each sub-plot), and of $\phi_{\text{peak}}$ and $\vartheta_{\boldsymbol{B}}$ (lower panel; blue and green shaded regions respectively), versus distance in arcminutes along cross-sections cutting through selected low-$p$ patches. The cross-sections are indicated and labeled on Figure \ref{fig:FornADepolRegions}. The units for $\phi_{\text{peak}}$ and $\vartheta_{\boldsymbol{B}}$ are rad m$^{-2}$ and rad [beam FWHM]$^{-1}$ respectively. The gray horizontal lines in the top panel indicate 25\%, 50\%, 75\%, 100\% of the maximum level of fractional polarization along the cross-section. The red shaded regions indicate the effective FWHM of the synthesized beam along the cross section, which we have centered on local minima in $p_{\text{FDS}}$.}
\label{fig:filamentaryMyDearWatson}
\end{figure*}

We investigated the magneto-ionized structure across the low-$p$ patches / $\phi_{\text{peak}}$ interfaces in more detail by extracting $p_{\text{FDS}}$, $\phi_{\text{peak}}$, and $\vartheta_{\boldsymbol{B}}$ along a number of cross-sections cutting through them. In each case, the data values were extracted at the nearest-neighbor pixel along the cross-section. The cross-sections are indicated on Figure \ref{fig:FornADepolRegions}, and the extracted data are plotted in Figure \ref{fig:filamentaryMyDearWatson}. We include data from six such cross-sections to supply a representative picture of the observed behaviors. Depending on the orientation of each cross-section, the effective FWHM of the synthesized beam varies between 21 and 27 arcseconds.

The plots firstly confirm that the low-$p$ patches are not merely the result of under-resolved monotonic changes in either Faraday depth or intrinsic polarization angle. The degree of depolarization expected in the former case can be calculated using Eqn. 2 of \citet{Laing2008}:

\begin{eqnarray}
p(\lambda^2) \approx p_{[\lambda=0]}e^{-2|\nabla RM|^2\sigma^2\lambda^4}
\label{Ch4eqn:LaingEqn}
\end{eqnarray}

\noindent where $p_{[\lambda=0]}$ is the intrinsic fractional polarization, $\nabla$RM is the gradient of the RM across the synthesized beam, and $\sigma^2$ is the standard width parameter for a circular Gaussian synthesized beam. At a weighted-mean-wavelength-squared value of 0.025 m$^2$ (i.e. relevant to the fractional polarization outputted from RM synthesis), and for Faraday depth gradients of 10, 20, and 50 rad m$^{-2}$ across the FWHM of a synthesized beam, we would expect to see depolarization from the maximum level observed in a nearby, off-patch position to $\sim$0.98, $\sim$0.91, and $\sim$0.56 of that maximum value, respectively. The plots of $p_\text{FDS}$ and $\phi_\text{peak}$ presented in Figure \ref{fig:filamentaryMyDearWatson} demonstrate that the level of observed depolarization along each cross-section is generally too large to be explained in this way. Similarly, the spatial gradient in the projected magnetic field orientation does not generally achieve the level of $\sim1$ radian per synthesized beam-width that would be required to produce significant depolarization. \\

Figures \ref{fig:FornADepolRegions} and \ref{fig:FornAFDRegions} reveal that complex spatial changes in $\phi_{\text{peak}}$ and $\vartheta_{\boldsymbol{B}}$ are associated with the low-$p$ patches, and therefore, the behavior of the $p_{\text{FDS}}$, $\phi_{\text{peak}}$, and $\vartheta_{\boldsymbol{B}}$ over the cross-sections will shed light on the physical processes operating there. Examination of the plots for cross-sections X1--X6 generally reveal that the interfaces in $\phi_{\text{peak}}$ (see Section \ref{sec:spatialFD}) appear to be approximately coincident with the point of minimum fractional polarization, and appear to form a spine around which broader low-$p$ complexes appear. At the $\phi_{\text{peak}}$ interfaces, the Faraday depth jumps by 50--100 rad m$^{-2}$ on average (and, we note, up to $\sim$100s of rad m$^{-2}$ when this experiment is repeated with the higher-resolution data presented in Figures \ref{fig:FornAFDZoomRegionsDP_paper} and \ref{fig:FornAFDZoomRegionsFD_paper}). At the distance of Fornax A, this implies the existence of a marked change in magnetoionic structure (and where the sign of $\phi_\text{peak}$ changes sign, \emph{magnetic} structure) over linear scales of less than a few kiloparsecs. More often than not, the absolute value of $\phi_{\text{peak}}$ appears to increase increasingly rapidly as the interfaces are approached, but this is not always the case --- see for example the $\phi_\text{peak}$ curves over cross-sections X4 and X6. 

\subsection{Spectropolarimetric modeling}\label{sec:FreqDepPol}

\begin{sidewaysfigure*}
\vspace{-4in}
\includegraphics[width=1.0\textwidth]{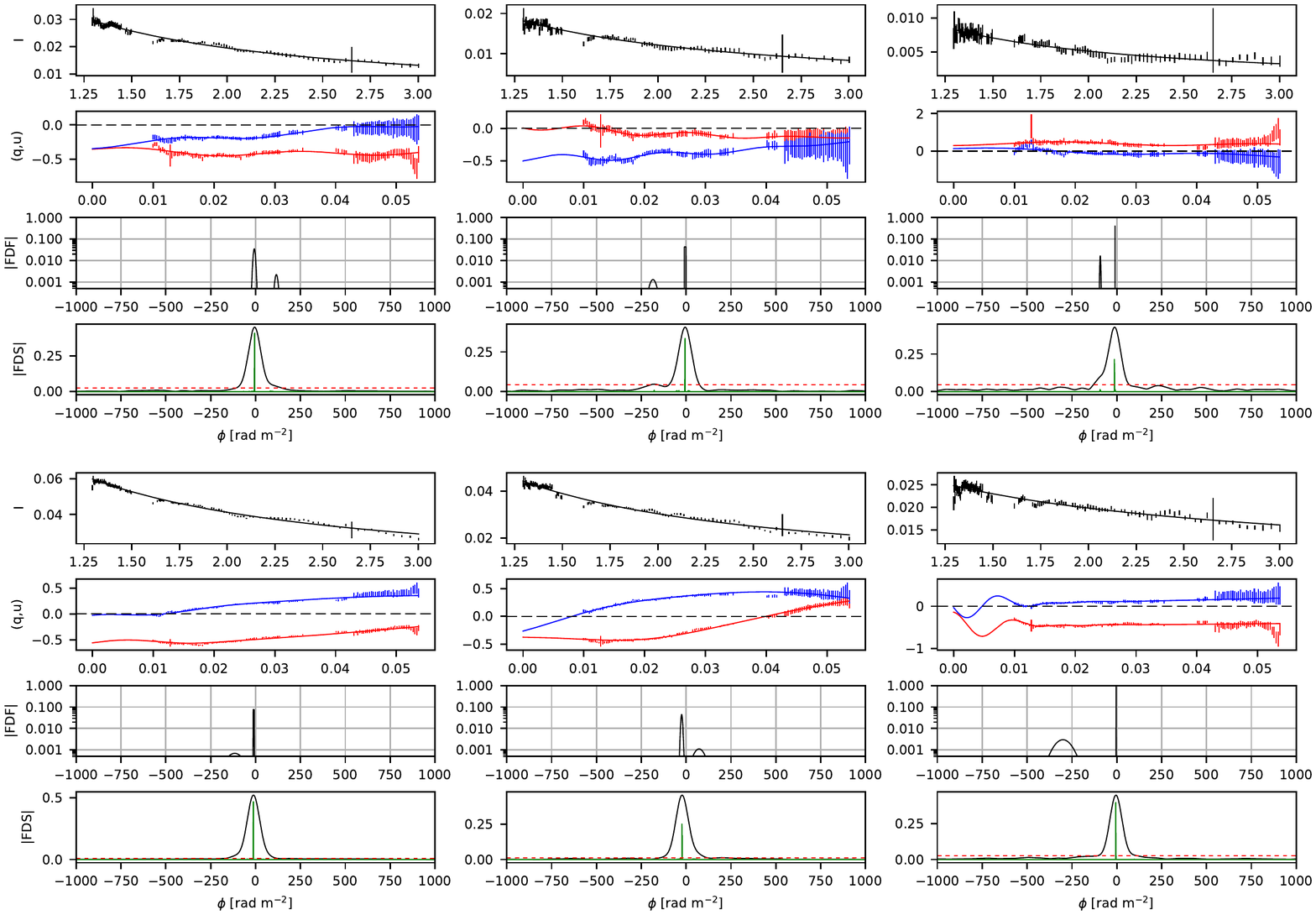}
\caption{Spectropolarimetric data and models for six selected pixel locations the lobe body (i.e. outside the low-$p$ patches; $p_\text{FDS}>0.4$). Data for the six pixel locations are laid out in $2\times3$ grid, with each location in that grid consisting of four vertically-tiled sub-panels. From top to bottom, these sub-panels show (respectively) the Stokes $I$ spectrum + power law model fit, the Stokes $q$ and $u$ spectra + best-fit model, the FDF corresponding to the best fit polarization model, and the FDS outputted from RM synthesis. The units on the x-axis of the four panels are respectively GHz, m$^2$, rad m$^{-2}$, and rad m$^{-2}$. The units on the y-axis of the four panels are respectively Jy, dimensionless, [rad m$^{-2}$]$^{-1}$, and beam$^{-1}$ [rad m$^{-2}$]$^{-1}$ RMSF$^{-1}$. The dashed black line in the ($q$,$u$) plot indicates zero polarized fraction. The red dashed line in the $|$FDS$|$ plot indicates our {\sc rmclean} cutoff level at 10$\sigma$ Gaussian equivalent significance (see main text). The green vertical lines in the $|$FDS$|$ plot indicate the position and amplitudes of {\sc rmclean} components. Note the logarithmic scale on the plots of the $|$FDF$|$ models, and that this differs from the plots of the reconstructed $|$FDS$|$. Corresponding numerical fitting results are presented in Table \ref{tab:fitgoodness}.}
\label{fig:RMSynthQUFits_filtype_body}
\end{sidewaysfigure*}

\begin{sidewaysfigure*}
\vspace{3in}
\centering
\includegraphics[width=1.0\textwidth]{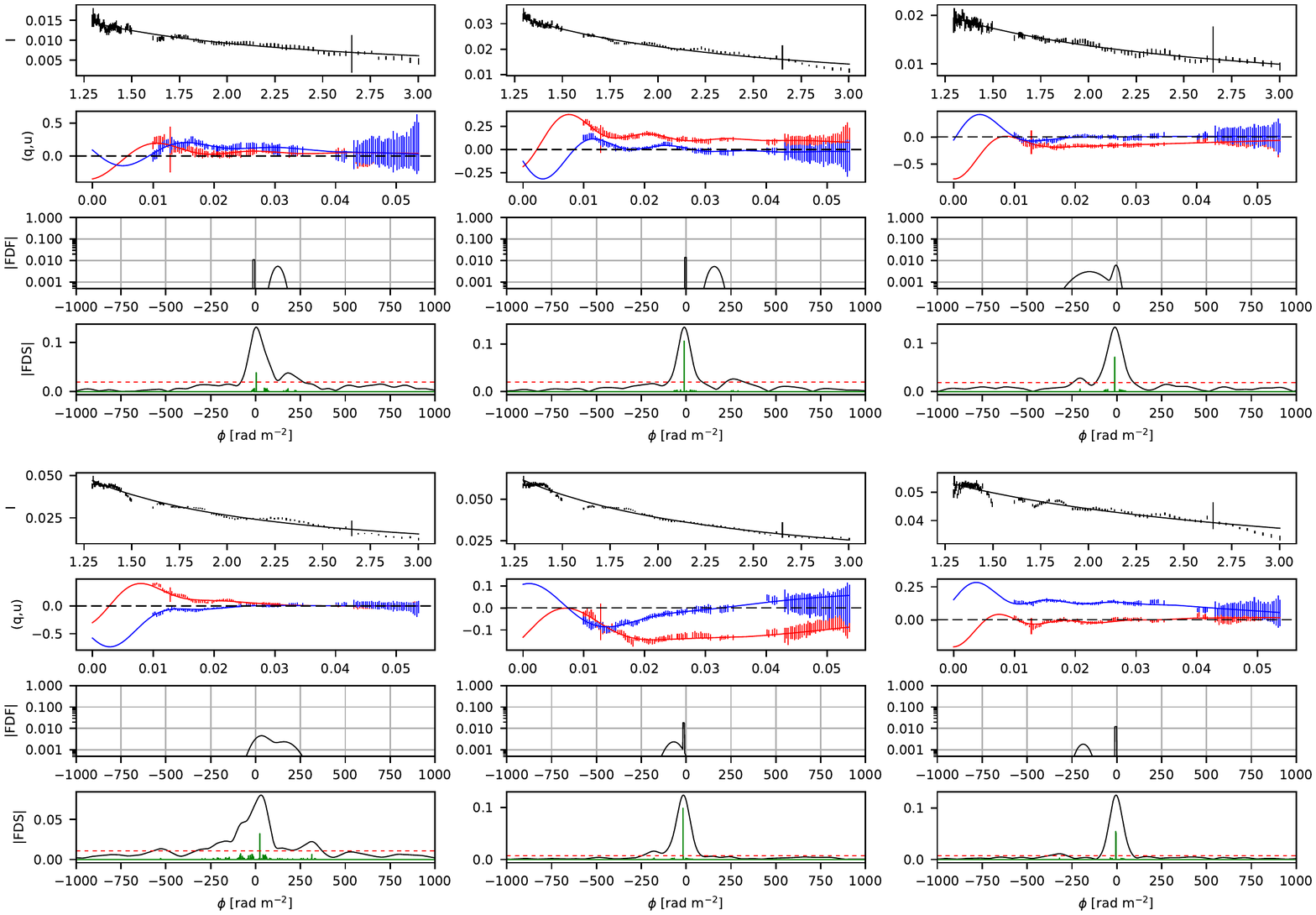}
\caption{As for Figure \ref{fig:RMSynthQUFits_filtype_body}, but for six selected pixel locations intermediate between the edge of the low-$p$ patches and their spine ($0.2>p_\text{FDS}\geq0.08$)}
\label{fig:RMSynthQUFits_filtype_inter}
\end{sidewaysfigure*}

\begin{sidewaysfigure*}
\vspace{-4in}
\centering
\includegraphics[width=1.0\textwidth]{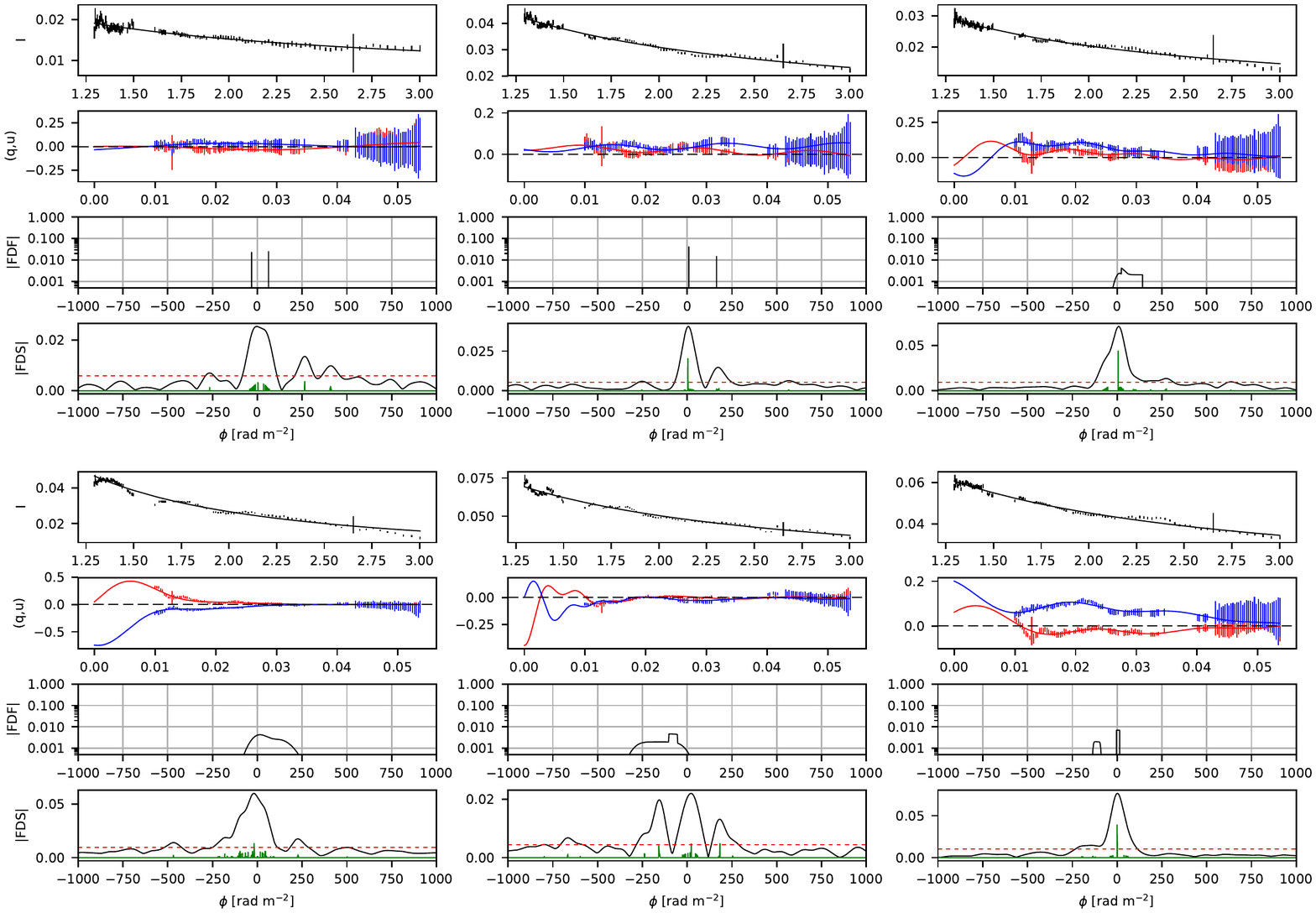}
\caption{As for Figure \ref{fig:RMSynthQUFits_filtype_body}, but for six selected pixel locations in the spine of the low-$p$ patches ($p_\text{FDS}<0.08$)}
\label{fig:RMSynthQUFits_filtype_spine}
\end{sidewaysfigure*}

\begin{figure*}
\centering
\includegraphics[width=\textwidth]{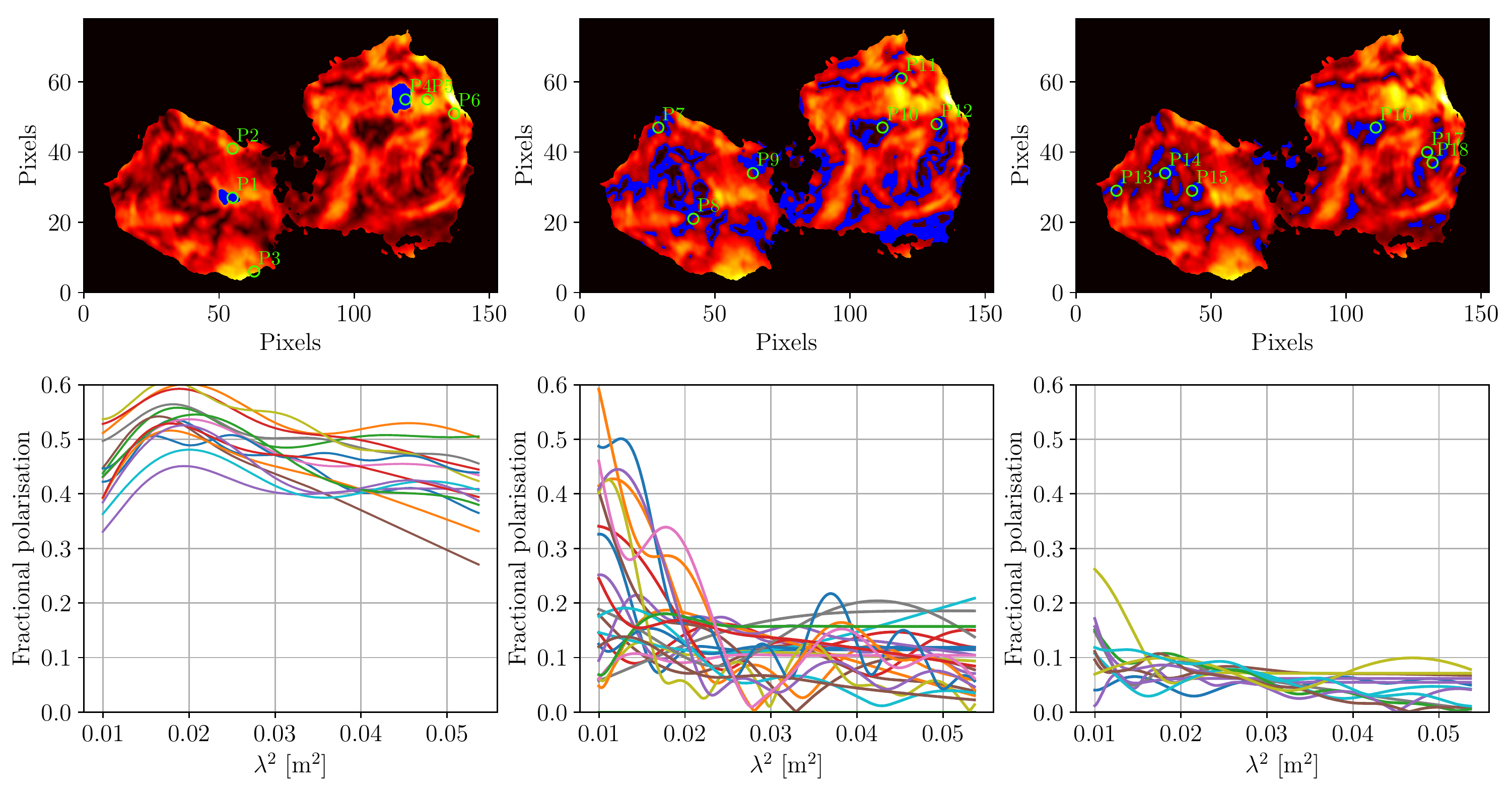}
\caption[]{$p(\lambda^2)$ behaviors associated with our best-fitting polarization models in the low-$p$ patches (see Section \ref{analysis}). The first, second, and third column of panels show maps of $p_{\text{FDS}}$ (upper row of panels) and plots of $p(\lambda^2)$ (lower row of panels) for pixels with $p_{\text{FDS}}\geq0.4$ (and with a further requirement that these pixels be near the central regions of the lobes), $0.2>p_{\text{FDS}}\geq0.08$, and $p_{\text{FDS}}<0.08$ respectively. These limits were chosen arbitrarily to correspond to `off-patch' locations in the lobe body, the and the periphery and central regions of the low-$p$ patches respectively. The pixels that satisfy these inequalities for each case are indicated in blue on the maps of $p_{\text{FDS}}$ in the upper row of panels. We remind the reader that one square pixel corresponds to one synthesized beam area in these spatially down-sampled images (see Section \ref{analysis}). In the lower row of panels, we plot $p(\lambda^2)$ curves for randomly selected pixel locations (i.e. beams) --- one in every four, 40, and 20 pixels locations from `off-patch', peripheral, and central regions of the low-$p$ patches respectively. Note that the green open circles and associated labels are unrelated to the $p(\lambda^2)$ plots displayed in this figure, but instead show the pixel locations selected for the plots show in Figs. \ref{fig:RMSynthQUFits_filtype_body}--\ref{fig:RMSynthQUFits_filtype_spine}.
}
\label{fig:depolModelPlotsFornAFilaments}
\end{figure*}

We now describe the results of our ($q$,$u$)-fitting analysis (Section \ref{analysis}). We start by presenting raw spectropolarimetric data and models for a sample of pixel locations.  We randomly selected eighteen such locations for this purpose, divided evenly between regions satisfying $p_{\text{FDS}}\geq0.4$, $0.2>p_{\text{FDS}}\geq0.08$, and $0.08>p_{\text{FDS}}\geq0$, which is designed to isolate pixels outside, at the periphery of, and inside the low-$p$ patches respectively. The spectropolarimetric data from each location are plotted in Figs. \ref{fig:RMSynthQUFits_filtype_body}--\ref{fig:RMSynthQUFits_filtype_spine}, while the pixel locations themselves are indicated on Figure \ref{fig:depolModelPlotsFornAFilaments} (top row). Each of Figures \ref{fig:RMSynthQUFits_filtype_body}--\ref{fig:RMSynthQUFits_filtype_spine} consists of a $2\times3$ (row $\times$ column) grid of four vertically-tiled sub-panels. Each location in the grid presented data for one pixel location, with the four panels presenting (from top to bottom) the Stokes $I$ spectrum plus best-fit model, the Stokes $q(\lambda^2)$ and $u(\lambda^2)$ data plus best-fit $(q,u)$ model, the FDF corresponding to this model (e.g. see \citealp{Anderson2016,OSullivan2017}), and the reconstructed FDS outputted from RM synthesis and {\sc rmclean}. \\

Qualitatively, for the central and peripheral locations of the low-$p$ patches, the plots reveal that:

\begin{itemize}
\item Stokes $q$ and $u$ generally show considerable non-sinusoidal frequency-dependent structure (see Figs. \ref{fig:RMSynthQUFits_filtype_inter} \& \ref{fig:RMSynthQUFits_filtype_spine}) 
\item Multiple emission components are generally required in/detected in the model FDF/reconstructed FDS. Typically, each component strongly influences the observed spectropolarimetric behavior.
\item The components often emit over a substantial range in Faraday depth, based on the widths of components required in the best-fit model FDF, and the location of {\sc rmclean} components in the reconstructed FDS
\end{itemize}

\noindent In contrast, outside the low-$p$ patches (see Figure \ref{fig:RMSynthQUFits_filtype_body}), it is evident that: 

\begin{itemize}
\item Stokes $q$ and $u$ remain greater in magnitude across the entire $\lambda^2$ range
\item While some degree of non-sinusoidal frequency-dependent behavior in Stokes $q$ and $u$ remains, the polarization behavior is generally dominated by single emission components with narrow $\phi$-widths and low absolute peak Faraday depths
\item Where additional components are present in the model FDF, they are comparatively narrow in $\phi$-space, and contribute a comparatively small amount of polarized flux
\end{itemize}

These observations generally extend to the full set of numerical results from ($q$,$u$)-fitting. The ($q$,$u$)-fitting results are recorded in Table \ref{tab:fitgoodness} for the pixel locations plotted in Figures \ref{fig:RMSynthQUFits_filtype_body}--\ref{fig:RMSynthQUFits_filtype_spine}. Columns 1-11 of the table contain (respectively): The pixel coordinate location of the fitted data, whether the pixel is located outside (denoted `O'), at the periphery of (P), or in the spine (S) of the low-$p$ patches, the fitted Stokes $I$ spectral index, the band-averaged Stokes $I$ flux density (denoted $<I>_\text{band}$), the best-fit parameter values and their uncertainties for the $j$-th emission component in the best-fit model for $p_{0}$, $\psi_{0}$, RM, $\sigma_{\text{RM}}$, and $\Delta \phi$, and finally, the reduced-$\chi^2$ (denoted $\tilde{\chi}^2$) and AIC values of the best-fit model for the pixel location concerned. We note that the type and number of emission components in the best-fit model can be inferred on the basis of blanked entries in the table. In a separate table (Table \ref{tab:fitsummary}), we record several statistics that provide a summary of the ($q$,$u$)-fitting results for all fitted pixel locations inside and outside the low$-p$ patches. Columns 1--8 of Table \ref{tab:fitsummary} contain (respectively): The type of region in relation to the low-$p$ patches (again, O, P, or S), the number of pixel locations (i.e. independent synthesized beam areas; see Section \ref{analysis}) analyzed in each such region, the proportion of pixel locations in each region that have: (i) a single model uniquely favored by our AIC criteria, (ii) external depolarization terms only in the best-fit model, (iii) internal depolarization terms only in the best-fit model, and (iv) both external and internal depolarization terms in the best-fit model, and finally, the mean values of the fractional-polarization-weighted values of $\Delta \phi$ and $\sigma_{\text{RM}}$ for the best-fit model over its $j$ emission components, which are defined as:

\begin{eqnarray}
\Delta\phi_{wtd}=\sum_{j}p_{0,j}\Delta\phi_j \bigg/  \sum_{j}p_{0,j}
\label{eqn:wtdDel}
\end{eqnarray}

\noindent and:

\begin{eqnarray}
\sigma_{\text{RM},wtd}=\sum_{j}p_{0,j}\sigma_{\text{RM},j} \bigg/  \sum_{j}p_{0,j}
\label{eqn:wtdSig}
\end{eqnarray}

\begin{deluxetable*}{c c c c c c c} 
\tiny
\tabcolsep=0.00cm
\tablecaption{($q$,$u$)-fitting results for data presented in Figures \ref{fig:RMSynthQUFits_filtype_body}--\ref{fig:RMSynthQUFits_filtype_spine} (see also Section \ref{analysis})} 
\tablehead{ 
\colhead{(1)} & \colhead{(2)} & \colhead{(3)} & \colhead{(4)} & \colhead{(5)} & \colhead{(6)} & \colhead{(7)}\\ 
\colhead{Pixel} & \colhead{Location} & \colhead{$\alpha$}  & \colhead{$<I>_\text{band}$} & \colhead{$p_{0[1,2,3]}$} & \colhead{$\psi_{0[1,2,3]}$} & \colhead{RM$_{[1,2,3]}$}\\ 
\colhead{ } & \colhead{ } & \colhead{ } & \colhead{[mJy beam$^{-1}]$} & \colhead{} & \colhead{[rad]} & \colhead{[rad m$^{-2}$]}
 }  \\
\startdata 
(55, 27) & O & -0.93 & 28 & 0.465(0.008),0.041(0.010),-(-) & -1.21(0.02),-1(2),-(-) & -6.9(0.8),110(30),-(-) \\					
(55, 41) & O & -0.83 & 15 & 0.46(0.01),0.06(0.03),-(-) & -0.75(0.02),-1.1(0.2),-(-) & -5.8(0.8),-190(10),-(-)\\ 				
(63, 6) & O & -1.19 & 9 & 0.39(0.03),0.1(0.2),-(-) & 0.18(0.10),-2(2),-(-) & -11(3),-100(300),-(-) \\ 							
(119, 55) & O & -0.76 & 52 & 0.564(0.008),0.2(0.1),-(-) & -1.45(0.04),-0.1(0.8),-(-) & -12(1),-0(100),-(-) \\ 					
(127, 55) & O & -0.69 & 37 & 0.47(0.02),0.1(0.2),-(-) & -1.34(0.07),-0.4(0.9),-(-) & -26(3),100(200),-(-) \\ 					
(137, 51) & O & -0.42 & 21 & 0.456(0.003),0.34(0.02),-(-) & 1.53(0.01),-1.2(0.3),-(-) & -3.3(0.5),-200(30),-(-) \\ 				
(29, 47) & P & -0.91 & 12 & 0.22(0.04),0.14(0.02),-(-) & -1.5(0.2),0.6(0.2),-(-) & 130(10),-5(6),-(-)\\ 						
(42, 21) & P & -0.86 & 29 & 0.15(0.01),0.29(0.09),0.12(0.06) & 0.1(0.1),-1.3(0.7),-0.3(0.8) & -5(4),150(80),-200(300) \\	
(64, 34) & P & -0.71 & 17 & 0.6(0.1),0.20(0.01),-(-) & -1(2),-1.46(0.09),-(-) & -160(40),-5(3),-(-) \\						
(112, 47) & P & -1.01 & 37 & 0.23(0.05),0.3(0.1),0.22(0.07) & 0.3(0.4),-1(2),-0.2(0.4) & -30(20),100(200),-40(90)\\			
(119, 61) & P & -0.82 & 50 & 0.164(0.008),0.3(0.3),-(-) & -1.09(0.08),1(2),-(-) & -15(3),-80(70),-(-) \\					
(132, 48) & P & -0.47 & 54 & 0.26(0.07),0.163(0.008),0.12(0.05) & -0.9(0.3),0.90(0.02),-1(1) & -30(50),-3.9(0.9),100(300) \\
(15, 29) & S & -0.55 & 18 & 0.034(0.004),0.2(0.3),-(-) & -0.2(0.2),1(1),-(-) & 59(10),200(300),-(-) \\	
(33, 34) & S & -0.64 & 38 & 0.047(0.006),0.7(0.3),0.02(0.02) & 0.3(0.2),-1.4(0.7),-1(1) & 10(6),150(70),200(300) \\			
(43, 29) & S & -0.81 & 26 & 0.12(0.01),0.27(0.03),-(-) & 0.24(0.07),-1.3(0.2),-(-) & 15(3),88(9),-(-) \\ 						
(111, 47) & S & -1.01 & 36 & 0.26(0.09),0.39(0.07),0.04(0.07) & -1.1(0.2),0.1(0.2),1.1(1.0) & 50(10),-30(20),-500(200) \\	
(130, 40) & S & -0.53 & 60 & 0.13(0.01),0.6(0.2),-(-) & -0.1(0.1),-0.9(0.4),-(-) & -78(4),-200(200),-(-) \\						
(132, 37) & S & -0.57 & 54 & 0.11(0.02),0.1(0.2),0.11(0.02) & 0.77(0.06),-0.5(0.6),1(1) & 6(3),-300(200),-10(40) \\				
\enddata 
\label{tab:fitgoodness} 
\end{deluxetable*} 

\addtocounter{table}{-1}
\begin{deluxetable*}{c c c c c c} 
\tiny
\tabcolsep=0.00cm
\tablecaption{\emph{...continued}} 
\tablehead{ 
\colhead{(1)} & \colhead{(2)} & \colhead{(8)} & \colhead{(9)} & \colhead{(10)} & \colhead{(11)}\\ 
\colhead{Pixel} & \colhead{Location} & \colhead{$\sigma_{\text{RM}[1,2,3]}$} & \colhead{$\Delta\phi_{0[1,2,3]}$} & \colhead{$\tilde{\chi}^2$} & \colhead{AIC}\\ 
\colhead{ } & \colhead{ } & \colhead{[rad m$^{-2}$]} & \colhead{[rad m$^{-2}$]} & \colhead{ } & \colhead{ }
 }  \\
\startdata 
(55, 27) & O & 5.4(0.7),8(7),-(-) & -(-),-(-),-(-) & 0.62 & -884 \\
(55, 41) & O & 9.6(0.9),19(9),-(-) & -(-),-(-),-(-) & 0.44 & -721 \\ 
(63, 6) & O & -(-),0(30),-(-) & -(-),-(-),-(-) & 0.46 & -342 \\ 
(119, 55) & O & 6(2),20(20),-(-) & 4(3),80(60),-(-) & 2.10 & -1117 \\ 
(127, 55) & O & 4(2),30(50),-(-) & -(-),-(-),-(-) & 0.81 & -1009 \\ 
(137, 51) & O & 0.1(0.9),71(2),-(-) & -(-),-(-),-(-) & 0.73 & -844 \\ 
(29, 47) & P & -(-),-(-),-(-) & 42(3),11(5),-(-) & 0.29 & -617 \\ 
(42, 21) & P & 7(4),20(30),100(300) & 7(4),60(10),0(300) & 0.71 & -962 \\ 
(64, 34) & P & 76(6),14(2),-(-) & -(-),-(-),-(-) & 0.47 & -763 \\ 
(112, 47) & P & 7(8),40(10),10(10) & 35(7),60(20),50(20) & 1.39 & -992 \\ 
(119, 61) & P & -(-),30(10),-(-) & 10(2),50(80),-(-) & 0.53 & -1202 \\ 
(132, 48) & P & 6(4),11(2),0(200) & 190(60),8(4),100(200) & 1.02 & -1234 \\ 
(15, 29) & S & -(-),0(200),-(-) & -(-),100(100),-(-) & 0.29 & -818 \\ 
(33, 34) & S & 2(5),80(40),0(300) & 9(5),90(30),0(200) & 0.43 & -1107 \\ 
(43, 29) & S &  20(3),-(-),-(-) & -(-),120(2),-(-) & 0.48 & -940 \\ 
(111, 47) & S & 10(10),40(6),0(200) & 90(10),10(10),0(100) & 0.78 & -1017 \\ 
(130, 40) & S & -(-),30(10),-(-) & 49.1(0.8),290(90),-(-) & 1.95 & -1188 \\ 
(132, 37) & S & 2(7),0(200),3(9) & 16(4),100(200),117(5) & 1.21 & -1214 
\enddata 
\label{tab:fitgoodness} 
\end{deluxetable*} 
\clearpage

\begin{deluxetable*}{c c c c c c c} 
\tablefontsize{\tiny} 
\setlength{\tabcolsep}{0.015in} 
\tablewidth{0pt} 
\tablecaption{Summary of ($q$,$u$)-fitting results for locations outside (O), at the periphery of (P), and in the spine of (S) the low-$p$ patches} 
\tablehead{ 
\colhead{(1)} & \colhead{(2)} & \colhead{(3)} & \colhead{(4)} & \colhead{(5)} & \colhead{(6)} & \colhead{(7)}\\ 
\colhead{Location} & \colhead{No. pixels} & \colhead{Proportion single} & \colhead{Proportion} & \colhead{Proportion} & \colhead{Mean} & \colhead{Mean}\\ 
\colhead{} & \colhead{} & \colhead{AIC-favored model} & \colhead{one/two/three comps.} & \colhead{external (only)/internal (only)/mixed} & \colhead{$\Delta\phi_{wtd}$} & \colhead{$\sigma_{\text{RM},wtd}$}\\ 
\colhead{} & \colhead{} & \colhead{best-fit model} & \colhead{in best-fit model} & \colhead{physics in best-fit model} & \colhead{} & \colhead{}\\ 
\colhead{} & \colhead{} & \colhead{} & \colhead{} & \colhead{} & \colhead{[rad m$^{-2}$]} & \colhead{[rad m$^{-2}$]}
 }  
\startdata 
O & 62 & 0.48 & 0.00, 0.66, 0.34 & 0.16, 0.15, 0.69 & 11 & 24\\ 
P & 961 & 0.21 & 0.00, 0.72, 0.28 & 0.12, 0.29, 0.59 & 37 & 61\\ 
S & 316 & 0.19 & 0.00, 0.76, 0.24 & 0.11, 0.24, 0.64 & 49 & 75
\enddata 
\label{tab:fitsummary} 
\end{deluxetable*} 

Overall, the best-fit models provide good fits to the data, with mean $\tilde{\chi}^2$ values in each region of (respectively) 1.61, 0.59, and 0.55. The higher-than-ideal $\tilde{\chi}^2$ values outside the low-$p$ patches appear to indicate that more emission components, or components described by a function other than Eqn. \ref{eqn:ShaneMod}, are required to fully encapsulate the behavior of the bright, high S/N polarized emission from these regions. The lower-than-ideal $\tilde{\chi}^2$ values for the peripheral and spine regions appear to arise as a result of a slight overestimation of the uncertainties on Stokes $q$ and $u$. This can be seen in Figs. \ref{fig:RMSynthQUFits_filtype_body}--\ref{fig:RMSynthQUFits_filtype_spine}, where the scatter between adjacent data points is noticeably lower than the amplitude of their error bars (particularly at high $\lambda^2$ values, where the frequency span of each $\lambda^2$ channel is small --- see Section \ref{imaging}). This does not affect our analysis or results in any important way --- the additional components in the model FDF are clearly required to fit large, systematic changes in Stokes $q$ and $u$, which persist over ranges in $\lambda^2$ space that are much larger than the inter-point separation. 

For essentially all fitted pixel locations, two or three component models are preferred by the AIC over single component models (column 4 of Table \ref{tab:fitsummary}). In fact, based on our AIC model selection criteria, single-component models represent a plausible alternative to the best-fitting two- and three-component models at only 6\% of pixel locations. Thus, the polarization behaviors in and around the low-$p$ patches are mostly inconsistent with simple Faraday rotation by a uniform magnetized thermal plasma in the foreground. Beyond this though, multiple different model types often have AIC values within 10 points of one another (column 3 of Table 2) --- for these pixel locations we cannot identify a unique best fit model with a high level of confidence. Inconveniently but unsurprisingly, this is more often the case in the spine and periphery of the low-$p$ patches than outside them, owing to the heavily reduced fractional polarization in these regions. Substantially more sensitive observations would be required to break these degeneracies. 

While we often fail to identify a single best-fitting model type, we can look for evidence that a particular type of depolarization physics (e.g. external Faraday dispersion, internal differential Faraday rotation) is uniquely associated with the low-$p$ patches. We took the model type with the lowest AIC value for each pixel (the most strongly favored best-fit model type, if not a uniquely favored best-fit model type), and classified it as having Faraday rotation only (the terms containing $\sigma_{\text{RM}}$ and $\Delta\phi$ in Eqn. \ref{eqn:ShaneMod} are equal to one for all emission components), external dispersion only (the term containing $\Delta\phi$ in Eqn. \ref{eqn:ShaneMod} is equal to one for all emission components), internal differential Faraday rotation only (the term containing $\sigma_{\text{RM}}$ in Eqn. \ref{eqn:ShaneMod} is equal to one for all emission components), or mixed (each of the $\sigma_{\text{RM}}$ and $\Delta\phi$ terms in Eqn. \ref{eqn:ShaneMod} are not equal to one for at least one emission component). We note that in reality, the generality of Eqn. \ref{eqn:ShaneMod} ensures that `mixed' models describe a wide range of polarimetric behaviors that do not strictly conform to the internal or external mechanisms. Thus, a `mixed' designation does not imply that the precise internal and external depolarization mechanisms modeled by the terms in this equation are necessarily in operation, either separately or together. With that said, we find no evidence to link specific types of depolarization physics to the low-$p$ patches (column 5 of Table 2). This strongly suggests that the magneto-ionized structure of the low-$p$ patches and their surrounds is complicated: The observed depolarization results from the combined effect of multiple interfering emission components, and various poorly-constrained depolarization mechanisms acting on this emission.  
 
Nevertheless, we can readily determine the characteristic range of Faraday depths and width in Faraday depth space that different components in the best-fit models emit over. The mean values of $\Delta\phi_{wtd}$ and $\sigma_{\text{RM},wtd}$ (see Eqns. \ref{eqn:wtdDel} and \ref{eqn:wtdSig}) over the analysed pixel locations are recorded in columns 6 and 7 of Table 2, broken down by region. In general, the locations outside the low-$p$ patches have comparatively low values of $\sigma_{\text{RM}}$ and $\Delta\phi$, where the fits are well-constrained by the data. The value of $\sigma_{\text{RM}}$ ranges up to 32 rad m $^{-2}$ with an average value of 24 rad m $^{-2}$, and a typical uncertainty of $<5$ rad m $^{-2}$. $\Delta\phi$ ranges up to 60 rad m $^{-2}$, with an average value of 11 rad m $^{-2}$, and a typical uncertainty of $<5$ rad m $^{-2}$. For the peripheral locations in the low-$p$ patches, $\sigma_{\text{RM}}$ ranges up to 90 rad m $^{-2}$ with an average value of 61 rad m $^{-2}$, and a typical uncertainty of $\sim5$ rad m $^{-2}$. The value of $\Delta\phi$ ranges up to 314 rad m $^{-2}$, with an average value of 37 rad m $^{-2}$, and a typical uncertainty of  $<10$ rad m $^{-2}$. For the pixel locations in the spine of the low-$p$ patches, $\sigma_{\text{RM}}$ ranges between 7 and 96 rad m $^{-2}$ with an average value of 75 rad m $^{-2}$, and a typical uncertainty of several 10s of rad m $^{-2}$. The value of $\Delta\phi$ ranges up to 330 rad m $^{-2}$, with an average value of 49 rad m $^{-2}$, and a typical uncertainty of $\sim$10 rad m $^{-2}$. \\

Having described our modeling results, we now attempt to isolate and display characteristic spectropolarimetric behaviors associated with the low-$p$ patches. To do so, we  once again distinguish between pixels located outside, at the periphery of, and inside the low-$p$ patches, according to the inequalities on $p(\lambda^2)$ previously discussed. We indicate each such region in blue on $p(\lambda^2)$ maps of the lobes in columns 1--3 of the upper row of panels in Figure \ref{fig:depolModelPlotsFornAFilaments} respectively. In the lower row of panels, we plot $p(\lambda^2)$ curves for the best fit polarization model at a set of pixel locations randomly drawn from the highlighted blue regions --- one in every four, 40, and 20 pixel locations outside the low-$p$ patches, at the periphery of the patches, and in the spine of the patches respectively. Moving from the left-most to right-most column (i.e from regions outside the patches to regions in their spine), the plots of $p(\lambda^2)$ reveal a clear progression of behavior from higher to lower band-averaged fractional polarization. As discussed above, this is associated with the presence of multiple emission components emitting over a range of Faraday depths, and thus the low-$p$ patches are generated by Faraday depolarization. The central low-$p$ patch locations exhibit the heaviest depolarization over the broadest frequency ranges, which is unsurprising. However, in many cases $p(\lambda^2)$ is not completely depolarized over the entire band, and at $\lambda^2\lesssim0.015$ m$^2$, there is some indication on an up-tick in fractional polarization towards shorter wavelengths for locations near the spines of the low-$p$ patches. We also note that for the peripheral locations, the best-fit $p(\lambda^2)$ models often depolarize over the band, from fractional polarization values similar to those seen outside the low-$p$ patches, to values more characteristic of those seen near their spines. Once again, this strongly suggests that it is Faraday depolarization operating in the low-$p$ patches.  

In general however, this depolarization behavior is not monotonic but oscillatory, meaning that we are not observing the action of a turbulent foreground screen \citep{Burn1966}, but rather the presence of mixed synchrotron-emitting and Faraday-rotating material, and/or interference effects generated by multiple emission components. Comparing the $p(\lambda^2)$ behavior in the peripheral and central regions of the low-$p$ patches, we see this oscillatory depolarization behavior in both cases. However, compared to that for the central regions, the $p(\lambda^2)$ data for the intermediate regions appear to have a greater amplitude of oscillation on average. \\

\thispagestyle{empty}

\thispagestyle{empty}

\thispagestyle{empty}

\thispagestyle{empty}

In summary, ($q$,$u$)-fitting reveals that:

\begin{enumerate}
\item Strong, oscillatory (as a function of $\lambda^2$) depolarization is often observed towards the low-$p$ patches. This is not consistent with pure Faraday rotation, or with frequency-independent depolarization. It implies that a Faraday-rotating plasma exists along the line of sight which possesses a comparatively complicated magneto-ionized structure.
\item It is rarely the case that $\sigma_{\text{RM}}$ is well-constrained to have a high value while $\Delta\phi$ is well-constrained to have a low value, or vice versa. Thus, in general, the $\boldsymbol{P}(\lambda^2)$ behaviors cannot be uniquely attributed to either external Faraday dispersion or internal differential Faraday rotation --- again, the depolarising medium must have rather more complex structure.
\item The amount and character of the depolarization changes in a systematic way moving from `off-patch' positions towards the central spine of the low-$p$ patches. Secondary emission components in the FDF/FDS increasingly dominate the spectropolarimetric behavior. The differential Faraday depths and dispersions of these components tend to increase from typically $\sim10$ rad m $^{-2}$ in the lobe body, to several tens of rad m $^{-2}$ in the peripheral regions, and somewhat higher still in the inner regions of the low-$p$ patches. 

\end{enumerate}

\section{Discussion}\label{discussion} 

\subsection{Basic properties of the Fornax A lobes}\label{sec-summaryofproperties}

Our discussion makes use of the following basic lobe properties, which are taken from the literature or measured from our data.

\subsubsection{Sizes}\label{sec-sizes}

NGC 1316, the host galaxy of Fornax A, lies at a distance of 18 Mpc \citep{Feldmeier2007,Stritzinger2010}, so an angle of one arcminute corresponds to a projected linear distance of 5.2 kpc. The Fornax A radio lobes range between 20 and 28 arcminutes in projected angular diameter (Figure \ref{fig:FornA_zoom}), corresponding to $\sim100$--150 kpc in projected linear extent. We estimate their combined volume to be $\sim 2\times10^{15}$ cubic parsecs. The largest angular scale (LAS) of the system is $\sim50$ arcminutes, corresponding to 260 kpc. The typical projected width of a low-$p$ patch (i.e. the peripheral and inner regions defined in Section \ref{sec:FreqDepPol}) is $\sim6$ kpc, whilst their typical lengths and spine-to-spine separation are $\sim30$--40 kpc and $\sim10$--20 kpc respectively (Figure \ref{fig:FornADepolRegions}).

\subsubsection{Age}\label{sec-age}

In the lower atmosphere of NGC 1316, \citet{Lanz2010} derive a buoyant rise velocity of $v_{\text{buoy}}\approx270$ km s$^{-1}$. From this, we estimate that the lobes are approximately $\text{LAS}/(2v_{\text{buoy}})=0.45$ Gyr old --- around the middle of the 0.1--1 Gyr range proposed by \citet{Ekers1983}. 

\subsubsection{Thermal plasma content and environment}\label{sec-tpcontent}

To successfully model X-ray emission from the Fornax A lobes, \citet{Seta2013} required an emission component from thermal plasma with an electron density $n_{e,\text{lobe}}=3\times10^{-4}f^{-1/2}$ cm$^{-3}$ and total mass of $9\times10^{9}f^{1/2}$ $M_\odot$, where the volume filling factor $f$ is constrained to be $1\times10^{-3}<f^{1/2}<1$ from the inferred cooling timescale. We assume this plasma exists and is dominated by ionized hydrogen and helium nuclei with a typical ICM abundance ratio of 9:1 (respectively), and thus a mass density of $1.17n_{e,\text{lobe}} m_\text{prot}$, where $m_\text{prot}$ is the mass of a proton. 

The Fornax cluster consists of a main component centered on the cD galaxy NGC 1399, and in-falling subcluster to the southwest centered on NGC 1316 \citep{Drinkwater2001}. The main cluster ICM has been studied in detail: It possesses a central thermal electron density of $\sim6\times10^{-4}$ cm$^{-3}$ (cluster contribution only), and $\sim1.7\times10^{11}$ $M_\odot$ of gas enclosed inside a 100 kpc radius \citep{Jones1997,Paolillo2002}. The subcluster ICM has not been studied in similar detail. \citet{Seta2013} implicitly derive $n_{e,\text{ICM}}\sim6\times10^{-5}$ cm$^{-3}$ around Fornax A, by extrapolating a beta-model of the main cluster atmosphere from X-ray measurements at cluster-centric radii of 35--280 kpc (made by \citealp{Jones1997}) to the  $\sim$1.5 Mpc main-cluster-centric radius of Fornax A. The uncertainty in the extrapolated value is compounded by the fact that the south-west subcluster is relatively massive in its own right. It contains a total mass of $\sim2-6\times10^{13}$ $M_\odot$ --- at least a third the mass of the main cluster --- with a probable gas fraction of $>6$\% \citep{Dvorkin2015}, and other indications that it is gas-rich, such as member galaxies experiencing a high ongoing rate of star formation \citep{Drinkwater2001}. Thus, we suggest the aforementioned estimate of the ICM density near Fornax A is probably too low.
 

\subsubsection{Magnetic field}\label{sec-magfield}

Previous studies have found that the average magnetic field strength in the lobe body ($B_{\text{av}}$) falls in the range 1--4 $\mu$G \citep{Kaneda1995,Tashiro2009,Seta2013}. Most recently, \citet{McKinley2015} derive a value of $B_{\text{av}}=2.6$ ($\pm0.3$) $\mu$G through precise simultaneous modeling of synchrotron radio emission and inverse Compton-scattered X-ray emission. 

\subsection{Summary of our key results}\label{sec-summary}

Any proposed explanation for the low-$p$ patches must account for, or at least be consistent with, the following key observations:

\begin{itemize}
\item That the low-$p$ patches are (a) spatially-resolved, (b) centered around enhancements in the value of $|\phi_{\text{peak}}|$ of typically up to $\sim$50 rad m$^{-2}$ (and sometimes greater), (c) centered around reversals in the sign of $\phi_{\text{peak}}$, and (d) located in regions where the sky-projected magnetic field orientation exhibits complicated eddy-like structure (though see our caveat in Section \ref{sec:spatialB}).
\item That complex polarization models are generally required to reproduce the detailed depolarization and repolarization behaviors observed towards the low-$p$ patches, including the need for multiple Faraday-rotated emission components, possessing differential Faraday depths and dispersions of typically $\sim$several 10s, and up to $\sim$100s, of rad m$^{-2}$ (Section \ref{sec:FreqDepPol})
\item That the low-$p$ patches exhibit very low values of fractional polarization, implying that the depolarising medium must either reside at, or extend to within a small distance of, the lobe surface (as opposed to being limited in extent to a small volume deep within the lobe; see Section \ref{sec:overviewpol}). 
\end{itemize}

\subsection{On the Faraday depth structure associated with the low-$p$ patches}\label{sec-FDstruct}

Arguably, the most remarkable aspect of our observations is the Faraday depth structure observed towards the low-$p$ patches --- specifically, that $\phi_{\text{peak}}$ generally reverses sign across the central spine of these regions, while $|\phi_{\text{peak}}|$ increases systematically as these interfaces are approached.

The sign-reversals observed in $\phi_{\text{peak}}$ are significant. Where they occur, the magnitude of the change in $\phi_{\text{peak}}$ is typically 50--100 rad m$^{-2}$ over linear scales of $\sim$kpc or less (Section \ref{sec:spatialFD}) --- much larger than the $\mathcal{O}(-6)$ rad m$^{-2}$ Galactic RM foreground contribution in this area. This implies that structure in the LOS-projected magnetic field is an important causal factor for the observed Faraday depth structure.

The magnitude of the enhancements in $\phi_{\text{peak}}$ are difficult to explain using the (lobe-averaged) values for $n_{e,\text{lobe}}$ (for $f=1$) and $B_{\text{av}}$ provided in Section \ref{sec-summaryofproperties}, which for a fully regular field, will generate only $|\phi_{\text{peak}}|\sim\mathcal{O}(5)$ rad m$^{-2}$ over the width of a typical low-$p$ patch. This is about an order of magnitude below that required. In principle, the lobe-averaged $n_e$ and $B$ values could generate the required amount of Faraday rotation over path lengths of $\sim100$ kpc, but this would require a coherent, regular magnetic field to exist throughout almost the full depth of the lobes. While this might plausibly be the case in parts of the eastern lobe (see Figure \ref{fig:FornA_B_flow}), the $|\phi_{\text{peak}}|$ enhancements and low-$p$ filaments are not located there, but rather in places where the projected field is knotted on smaller scales (Section \ref{sec-summary}). Thus, it must be that either (a) $n_{e}$, $B$, or both in combination are a factor of 10--$20\times$ higher than the lobe-averaged values towards the low-$p$ patches, assuming that Faraday rotation occurs over a path length equal to the typical width of a low-$p$ patch, or if we relax this last assumption, that (b) local increases in $n_{e}$, $B$, and the magnetic field coherence length all act together to produce the required Faraday depths/dispersions.

In either case, we conclude that the low-$p$ patches and their associated Faraday depth structure must be generated by regions in the lobes where the local electron density and/or magnetic field strength is elevated above the lobe average(s). We note that \citet{McKinley2015} and \citet{Ackermann2016} reached a similar and potentially related conclusion --- namely, that the Fornax A lobes must contain relatively dense thermal plasma agglomerations, in order to successfully explain details of the (Stokes $I$) radio, X-ray, and $\gamma$-ray emission from the lobes. We discuss the possible physical nature of this magnetized substructure in Section \ref{sec-physscenario}.

\subsection{Mass, location, and source of the thermal plasma associated with the low-$p$ patches}\label{sec-reqmass}

We can estimate the minimum mass of thermal plasma associated with the low-$p$ patches in a general way, as follows. The peak Faraday depth of a column of thermal electrons threaded by a uniform magnetic field is (e.g. \citealt{HH2012,Anderson2015}):

\begin{eqnarray}
\phi &=& 0.81n_eB_{u,||}L \nonumber \\
&=& 26N_{e,20}B_{u,||} ~\text{rad m}^{-2}
\label{eq:patchFD}  
\end{eqnarray}

\noindent where $B_{u,||}$ is the strength of the magnetic field projected along the line of sight [$\mu$G], and $N_{e,20}$ is the electron column density in units of $10^{20}$ cm$^{-2}$. Taking the $p_\text{FDS}$ map presented in Figure \ref{fig:FornADepolRegions}, then masking out regions where $p>0.1$ and which fall outside boxes A--G, I--M, O, and P, the solid angle occupied by the low-$p$ patches is $\sim4.1\times 10^5$ square arcseconds. This corresponds to a projected area $A_{\text{low-}p}=3.24\times10^{46}$ cm$^{2}$ at the distance of the lobes. Assuming the conversion factor between thermal electron density and plasma mass density given in Section \ref{sec-tpcontent}, the total mass of material in the patches is roughly:

\begin{eqnarray}
M_\text{Thermal}&\approx&1.17\times10^{20}N_{e,20}A_{\text{low-}p}m_\text{prot}
\label{eq:patchFD2}  
\end{eqnarray}

\noindent Assuming the low-$p$ patches are characterized a typical Faraday depth $\phi_{\text{low-}p}$, then combining Eqns. \ref{eq:patchFD} and \ref{eq:patchFD2} after solving for $N_{e,20}$, yields:

\begin{eqnarray}
M_\text{Thermal}&\approx&1.23\times10^8\bigg(\frac{\phi_{\text{low-}p}}{B_{u,||}}\bigg) ~M_\odot
\label{eq:patchFD3}  
\end{eqnarray}

\noindent from which we get $M_\text{Thermal}\approx1.2\times10^9~M_\odot$, using $\phi_{\text{low-}p}=25$ rad m$^{-2}$ (Section \ref{sec:spatialFD}) and $B_{u,||}=B_\text{av}$ (though our conclusions from Section \ref{sec-FDstruct} should be kept in mind here). In fact, the true mass is likely to be higher than this, since our mask will tend to exclude material lying at the far side of the lobes, and the calculation assumes the magnetic field is oriented along the LOS with no reversals. On the other hand, if the enhancements in $|\phi_\text{peak}|$ are primarily related to local enhancements in the magnetic field strength (Section \ref{sec-FDstruct}), the required plasma mass is reduced in proportion. Nevertheless, we can compare the value as stated with existing observational results in this area. Interestingly, if we split our calculated mass evenly between the two lobes, it agrees with the Seta et al. estimate for the western lobe if $f\approx0.004$, which is consistent with the limits that they place on this parameter (see Section \ref{sec-tpcontent}). If we note the projected area occupied by the low-$p$ patches and model them as cylindrical regions viewed in projection, their associated volume is $\sim1.7\times 10^{13}$ cubic parsecs, yielding $f\approx0.008$, which is obviously comparable. Thus we suggest that the thermal material detected by Seta et al. may in fact be associated with the low-$p$ patches. Future high-resolution X-ray observations can confirm or refute this. Moving beyond Fornax A, similar thermal plasma masses are found to be associated with the buoyant lobes of other radio galaxies --- e.g. \citet{Simionescu2009,Kirkpatrick2009,Salome2011,Werner2011,MN2012,OSullivan2013b,Russell2017}. 

Where in the system is this material located, and from where does it originate? It must be associated with the lobes themselves, or we could not explain the observed relationship between the low-$p$ patches and structure in the projected magnetic field orientation and Stokes $I$ filaments (Section \ref{sec:spatialB}). Direct entrainment of the ISM/ICM in the radio jet cannot provide the required mass transport rate \citep{LB2002,Seta2013}. One possibility suggested by simulations is that radio bubbles lift and mix thermal plasma into themselves through a combination of buoyant motion and large-scale Rayleigh-Taylor (R-T) instabilities (e.g. \citealp{Churazov2001,Reynolds2002,Pope2010,Weinberger2017}). The advected plasma masses therein typically range up to $10^9$ M$_\odot$ per lobe. This may be capable of satisfying our requirements, though we note that more material can be probably be lifted through multiple episodes of AGN activity (e.g. \citealp{OSullivan2013b,Russell2017}), by magnetic fields that increase coupling between gas phases in and around the lobes (e.g. \citealp{Russell2017}), or by additional hydrodynamic forces generated by certain buoyant flow behaviors \citep{Pavlovski2007}. For this scenario, an open question is whether NGC 1316 contained sufficient thermal mass to raise into the lobes to begin with --- \citet{Forman1985} and \citet{Kim1998} find that the total mass of gas in the host is now just $\sim1-3\times10^9$ $M_\odot$. However, the system has also experienced significant merger and interaction events since the formation of the lobes \citep{MF1998,Horellou2001,Beletsky2011}, which may have stripped gas from the host after the lobes formed.

While we favor the advection scenario just described, a plausible alternative is that the ICM previously-displaced by the lobes has begun to intrude back into them via Rayleigh-Taylor (R-T) instabilities \citep{Alexander2002,PS2006,SG2007b,SG2007}. Indeed, \citet{Seta2013} have already proposed that this might be the origin of the thermal material in the western lobe. However, their estimate of the thermal plasma density in the lobes is a factor of five greater (more, if $f<1$) than their estimate of that in the surrounding ICM (Section \ref{sec-tpcontent}). For the R-T mechanism to be viable then, the ICM must cool and condense significantly after penetrating the lobes, or otherwise, the true ICM density around Fornax A must be higher than they project, as we have already discussed (Section \ref{sec-tpcontent}). We explore the viability of the R-T scenario further in Section \ref{R-Tinstab}.

Finally, we note that the majority of low-$p$ patches appear to sit in a curved lane running through the system, extending from region P in the western lobe, through regions O, M, L, and I, and from there into the eastern lobe through regions G, F, E, D, C, B, and A. This feature is striking, and motivated us to consider whether a more exotic and inherently anisotropic process may have transported the thermal material into the lobe interior, such as the Bondi-Hoyle wake or stripped ISM of an external orbiting galaxy, or jet-induced mixing at the lobe-ICM boundary (e.g. \citealp{Wilson1989,NB1993}). Both of these possibilities seem unlikely: In the former case, because there are no sufficiently massive galaxies nearby, while in the latter, because there is no evidence of a strong jet having recently operated in the system (Section \ref{sec-intro}). Instead, we suggest that the arrangement of depolarising material is more likely to have been generated by vortical flow behaviors in the lobe, combined with the intrinsic axisymmetry of the system (e.g. \citealp{BK2001}). This idea is related to the model that we propose in Section \ref{sec-physscenario}.

\subsection{The mechanism that produces the low-$p$ patches and associated Faraday depth structure}\label{sec-physscenario}

The most significant challenge posed by our results is that of explaining all of the observed interrelationships between the low-$p$ patches, enhancements in the magnitude of the peak Faraday depths and associated sign-changes in such, the complex frequency-dependence of the linear polarization signal, and structure in the sky-projected magnetic field orientation (summarized in Section \ref{sec-summary}). We now evaluate several physical mechanisms that represent, in our view, the most plausible means of producing the required couplings.

\subsubsection{Kelvin-Helmholtz instabilities}\label{K-Hinstab}

For buoyant radio lobes rising through a surrounding ICM, a velocity shear will be set up across the lobe-ICM interface, and Kelvin-Helmholtz (K-H) instabilities may form. Considering the observational implications of a magnetic field embedded in these flow structures, \citet{Bicknell1990} proposed that this mechanism may give rise to the RM/depolarization bands observed towards some radio lobes. Modern generic simulations of the K-H instability confirm the basic plausibility of this connection, while somewhat modifying the underlying physical picture. In summary, they show that regular, coherent magnetic field structures can be pulled into the base of K-H vortices, while turbulence and magnetic reconnection events act to form a highly complex magnetized medium inside the vortex itself (\citealp{Obergaulinger2010,Karimabadi2013,Ma2014} and refs. therein). 

To date, simulations of lobe-ICM interfaces specifically have lacked the numerical resolution to track dynamically important microphysical effects in the vortex region \citep{Karimabadi2013}. Nevertheless, they suggest that this interface can become K-H unstable (e.g. \citealp{HE2011,Weinberger2017}), including in poor cluster environments \citep{HK2014,EHK2016}, but that this is sensitively dependent on factors such as the current state of jet operation (e.g. \citealp{PS2006,Sternberg2008}), the magnitude of viscous damping effects (e.g. \citealp{Reynolds2005,Kaiser2005,Roediger2013}), and the detailed magnetic field configuration of the lobes and that of their environment (e.g. \citealp{Kaiser2005,KA2007,Ruszkowski2007,DF2008,DS2009,EHK2016}).

We are unaware of simulations that are currently capable of generating detailed global maps of Faraday depth and depolarization arising from K-H instabilities in realistic lobe environments. Nevertheless, if such instabilities did form at the lobe-ICM interface, then broadly following \citet{Bicknell1990}, we suggest that:

\begin{enumerate}
\item In regions where the crest of a K-H wave is viewed from above (i.e. the part of the flow feeding into the vortex), the direction of $B_{||}$ will show a rapid reversal.
\item A complex mixture of Faraday-active material will form in the K-H vortex, potentially Faraday-depolarising emission from the background. Such regions will be spatially correlated with the regions described in the previous point.
\item Waves should appear on the lobe edges as well as on the face, and in the former case, the characteristic projected scale height of the resultant $\phi_{\text{peak}}$ depolarization structure should be $\sim$one-third of the inter-low-$p$-patch distance on the lobes face \citep{Bicknell1990}.
\end{enumerate} 

\begin{figure}
\centering
\includegraphics[width=0.40\textwidth]{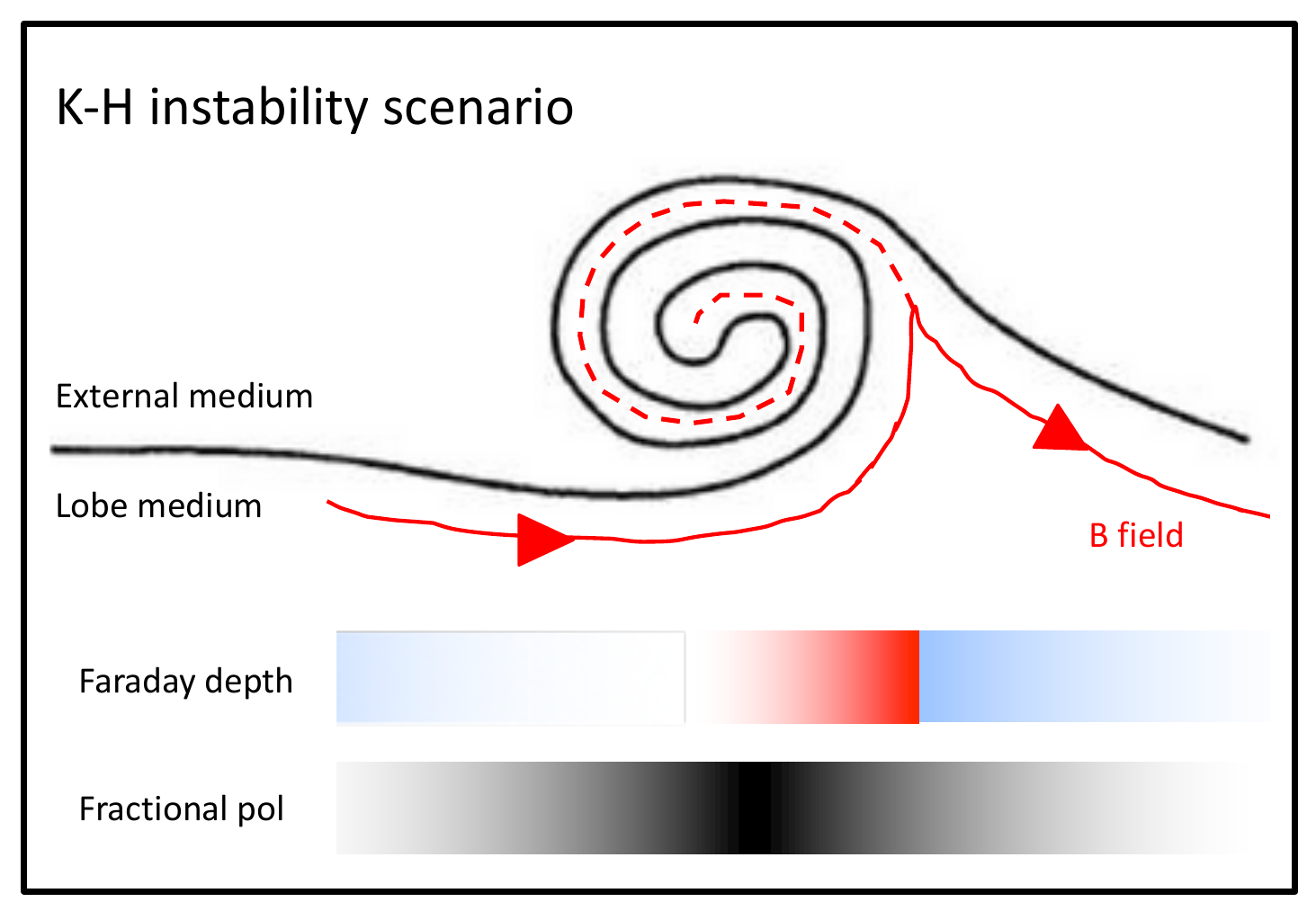}
\caption[Cartoon of a moderately well-developed K-H instability operating at a lobe-ICM interface]{Cartoon of moderately well-developed K-H instability operating at a lobe-ICM interface. The black solid line indicates the lobe-ICM interface. The red solid line indicates the magnetic filed in the lobe. We note that the external medium may also be magnetized, but omit this for clarity. The red dashed line indicates the location of magnetic field structure that is expected to be significantly affected by complicated MHD effects. The red/blue bar below the cartoon indicates where an observer looking down on the structure from above would see positive/negative Faraday depths (respectively, analogous to the color map in Figure \ref{fig:FornAFDRegions}) as a result of the coherent magnetic field structure entrained in the flow pattern. The grayscale bar indicates the level of fractional polarization seen by the observer (darker corresponds to lower fractional polarization), as determined by Faraday depolarization generated by turbulent/small-scale magneto-ionic structure in the K-H vortex. In this scenario, we propose that the observed interfaces in peak Faraday depth are generated by a reversal of the coherent LOS magnetic field in either the lobe or external medium, as it is pulled into the base of the K-H vortex. The associated depolarized patches are generated by the K-H vortex itself, in which complicated small-scale magnetoionic structure is expected to form --- e.g see \citet{Karimabadi2013}.}
\label{fig:KHinstab}
\end{figure}

\noindent The proposed scenario is depicted in Figure \ref{fig:KHinstab}. We suggest that the sign-reversals in $\phi_{\text{peak}}$ (Section \ref{sec:spatialFD}), the spatial coupling between such reversals and the low-$p$ filaments (Section \ref{sec:spatialFD}), and the $\phi_{\text{peak}}$/depolarization structure observed in boxes A, N, and R of Figures \ref{fig:FornADepolRegions} and \ref{fig:FornAFDRegions}), are qualitatively consistent with this scenario. Moreover, magnetic fields are naturally amplified in K-H vortices, providing a possible means through which substantial Faraday depths can be generated over relatively short path lengths.

The crux of the issue is whether the magnetic fields at the lobe-ICM interface are simultaneously strong enough to generate the required Faraday depths, but weak enough to render the lobe-ICM interface vulnerable to the instability. The former point has been discussed in Section \ref{sec-FDstruct}, noting that the scale-height of a K-H wave will be similar to the widths of the low-$p$ patches (as can be deduced from the scale-lengths provided in Section \ref{sec-sizes}, and the relationship between the wavelength and heights of K-H waves provided by \citealt{Bicknell1990}).

Addressing the second point, linear stability analysis shows that lobe-ICM interface is stable against K-H instabilities when the Alfv\'{e}n number (i.e. the ratio of the shear-to-Alfv\'{e}n velocities) satisfies (e.g. \citealp{Chandrasekhar1961,Obergaulinger2010}):

\begin{eqnarray}
\mathcal{A}\equiv\frac{\Delta v}{\sqrt{|\boldsymbol{B}|^2/\rho}}<2
\label{BstabCrit}
\end{eqnarray}

\noindent where $\Delta v$ is the shear velocity between the two flows, and $\rho$ is the mass density of the lighter of the two mediums. We use $\Delta v=v_{\text{buoy}}=2.7\times10^7$ cm s$^{-1}$ (Section \ref{sec-summaryofproperties}), implicitly assuming that the lobes act like a vortex ring, continuing to exhibit vortical flow even if they have ceased to rise buoyantly (e.g. see \citealp{Churazov2001}). The large-scale `swirling' morphology of the lobes would seem to support this. For values of $B_\text{av}$, $n_e$ ($f=1$), and the assumed plasma composition taken from Section \ref{sec-summaryofproperties}, we have $\mathcal{A}\sim0.25$ --- i.e. the interface is K-H stable.

To achieve instability (at least initially, and prior to the onset of non-linear stabilization/disruption effects --- see \citealp{Frank1996,Obergaulinger2010}), $\Delta v$ or $\rho$ must be higher by factors of eight and 64 respectively, or $B$ a factor of eight lower. A factor-of-eight increase in $\Delta v$ is not plausible, not least because this would require the lobe to be moving supersonically through the ICM (e.g. see \citealp{Machacek2005a}). The second possibility, and the third possibility in combination with our prior constraint on the value of the product $n_eB$, imply that $n_e\gtrsim0.02$ cm$^{-3}$ and $n_e\gtrsim0.05$ cm$^{-3}$ in the K-H-active layer respectively. In the diffuse Fornax cluster gas though, such high plasma densities are found only in the central arcminute of NGC 1399 --- the cD galaxy of the main cluster. Thus the K-H scenario seems unlikely, even in the absence of other difficulties, such as its apparent inability to explain the spatial coupling between the projected magnetic field structure of the lobes and the $\phi_\text{peak}$ and low-$p$ patches.

\subsubsection{Rayleigh-Taylor instabilities}\label{R-Tinstab}

Studies by \citet{Alexander2002,PS2006,BK2001,BK2002,Reynolds2002,DeYoung2003,SG2007b,SG2007} demonstrate that magnetized buoyant radio bubbles can become susceptible to R-T instabilities around $0.1$--$3\times10^8$ years after their formation, where `fingers' of the ICM intrude into the lobes after having been excluded at earlier times. For typical conditions, the fastest-growing unstable modes possess wavelengths on the order of $\sim10$ kpc --- similar to the characteristic length-scales of the low-$p$ patches and $|\phi_\text{peak}|$ structure that we seek to explain. If we extend this scenario in a basic way by speculating that:

\begin{enumerate}
\item the intruding ICM fingers carry a coherent magnetic field along with them
\item the fingers additionally contain turbulent magneto-ionic sub-structure, perhaps due to the formation of secondary K-H instabilities at the finger-lobe material interface
\end{enumerate} 

\begin{figure}
\centering
\includegraphics[width=0.40\textwidth]{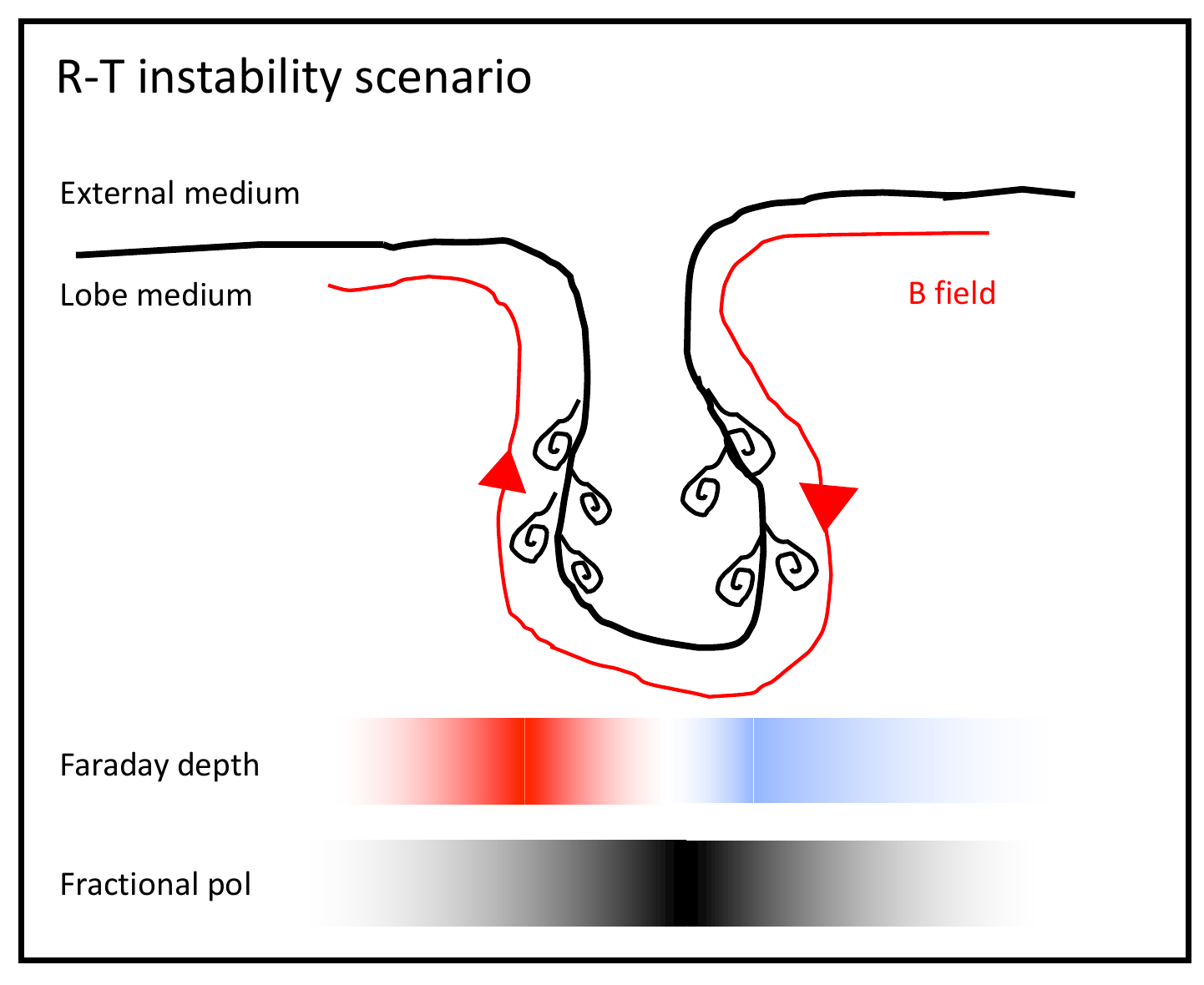}
\caption[Cartoon of moderately well-developed R-T and K-H instabilities operating at a lobe-ICM interface]{As for Figure \ref{fig:KHinstab}, but for the R-T instability scenario. The spirals on the side of the main finger indicate possible secondary K-H instabilities. In this scenario, we propose that the observed interfaces in peak Faraday depth are generated by the coherent magnetic field in either the lobe or external medium being pulled along with the intruding R-T finger. The associated depolarized patches are generated by complex magneto-ionic structure resulting from secondary K-H eddies at the finger boundary, or by unresolved changes in the magnetoionic structure of the finger itself.}
\label{fig:RTinstab}
\end{figure}

\noindent then we might plausibly observe a rapid reversal in the direction of $B_{||}$ across the tip of the finger when viewed from above, with accompanying depolarization (see Figure \ref{fig:RTinstab} for a schematic). Indeed, \citet{Seta2013} have already proposed that the presence of thermal plasma in the western lobe might be explained via this mechanism, so we start by assessing whether the R-T mechanism might operate, before commenting on its ability to explain our broader observations.

For the instability to go, the lobes must be at least slightly over-pressured relative to the external environment, which they may be by up to a factor of five (\citealp{Seta2013}, though with significant uncertainty; see Section \ref{sec-tpcontent}). \citet{Chandrasekhar1961} showed that the R-T instability is stabilized for all angular wavenumbers larger than $k_{\text{crit}}$, where

\begin{eqnarray}
k_{\text{crit}} = \frac{2\pi a(\rho_e-\rho_l)}{B^2\text{cos}^2\theta}
\label{RTcrit}
\end{eqnarray}

and $a$ is the acceleration of the interface, $\rho_e$ and $\rho_l$ are the mass densities of the external and lobe-based plasmas, $B$ is the magnetic field strength, and $\theta$ is the angle between the magnetic field and the wavevector associated with $k$. 

We do not know the acceleration of the lobe ICM-interface due to their pressure mismatch, but we can estimate the \emph{required} acceleration as follows. The R-T instabilities must operate on scales at least as small as 5 kpc, which corresponds to $k_c\approx4\times10^{-22}$ cm$^{-1}$. Rearranging Eqn. \ref{RTcrit} to solve for $a$, then setting $\rho_e=1.174\times10^{-28}$ g cm$^{-3}$ (based on the \citealt{Seta2013} estimate provided in Section \ref{sec-tpcontent}, which is uncertain, but see below), $\rho_l=0$ g cm$^{-3}$ (based on the assumption that the lobe contains no thermal plasma prior to the onset of the instability), $\theta=0$, and using our previous value for $B$, we find $a\approx4\times10^{-6}$ cm s$^{-2}$. The timescale for the instability to develop is then given by \citep{Chandrasekhar1961,DeYoung2003}:

 \begin{eqnarray}
t_{\text{R-T}} & = & \bigg\{\frac{ak_c}{\sqrt{3}(\rho_e+\rho_l)}\bigg[(\rho_e-\rho_l) - \frac{k_cB^2\text{cos}^2\theta}{2\sqrt{3}\pi a}\bigg]\bigg\}^{-1/2}\\
& \approx & 1.6 ~\text{Myr} \nonumber
\label{RTgrowth}
\end{eqnarray}
 
\noindent where all symbols have their previously defined meanings. This is the $e$-folding time, so several times this value is the approximate timescale at which the instability becomes well-developed. Note that the result is not especially sensitive to the value of $\rho_e$.

Thus, the lobe-ICM interface may be R-T unstable, and R-T fingers could form rapidly if this is so. However, a direct connection between R-T fingers and the observed Faraday depth structure seems unlikely for the following reasons: Given the estimated ICM density in the vicinity of the lobes, R-T fingers cannot generate the required Faraday depths over plausible path lengths without significant cooling or strong local enhancements in the local magnetic field strength (Section \ref{sec-FDstruct}). In the former case, it is not clear what mechanism would lead the ICM to cool suddenly upon intrusion into the lobes. The latter possibility will have the effect of inhibiting the formation of turbulent magnetoionic sub-structure in the R-T finger, which is necessary to produce Faraday depolarization in this scenario \citep{SG2007b,SG2007}.

\subsubsection{Advected ISM from NGC 1316}\label{Advect}

\begin{figure}
\centering
\includegraphics[width=0.40\textwidth]{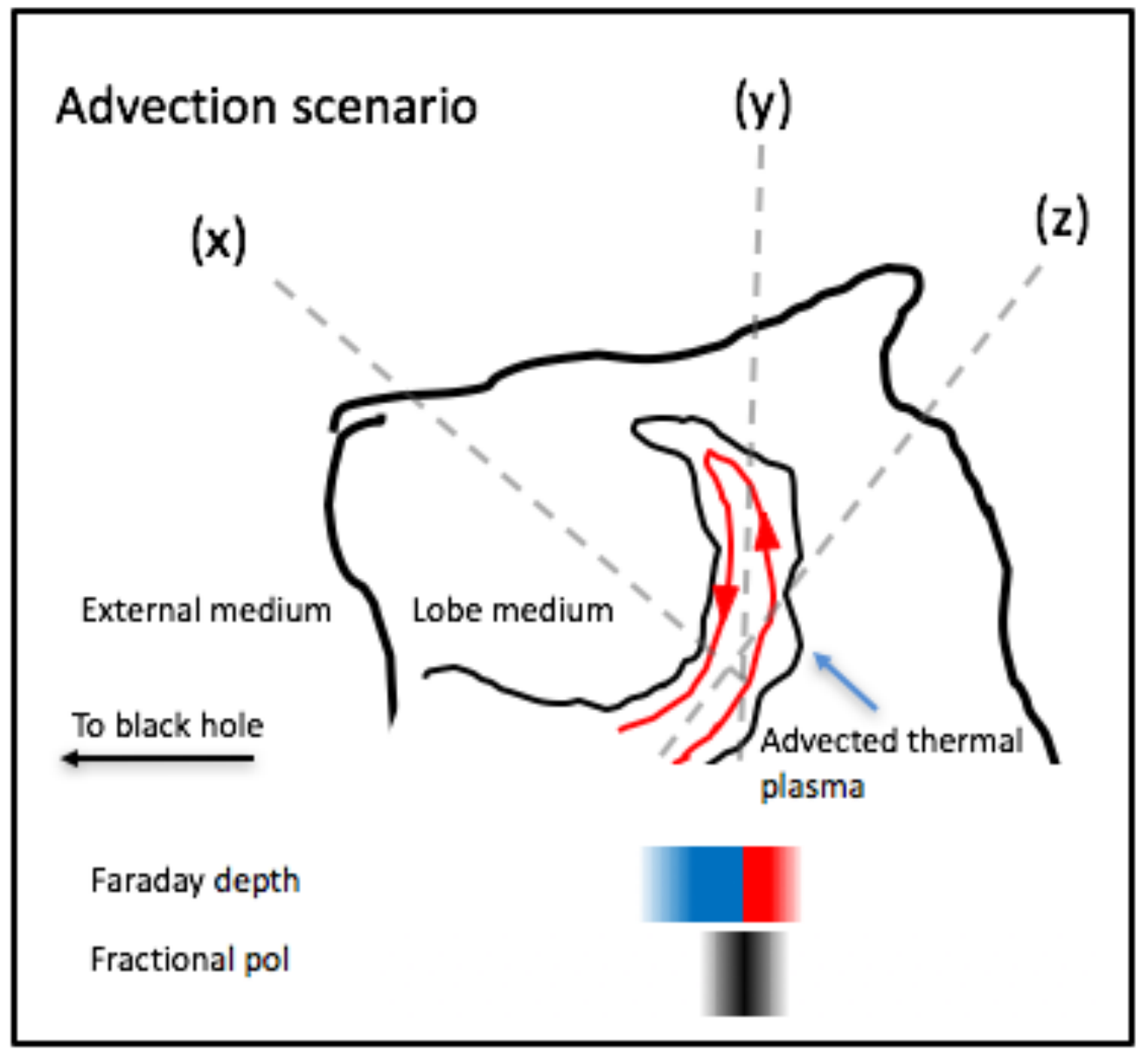}
\caption[]{As for Figure \ref{fig:KHinstab}, but for the thermal plasma advection scenario. In this scenario, we propose that the observed interfaces in peak Faraday depth are generated by the coherent magnetic field dragged along by thermal plasma inclusions in the lobes, which have been advected from the ISM of the host galaxy. The associated depolarized patches are generated by magnetoionic sub-structure in the advected material. The positions (x), (y) and (z) represent different lines of sight (gray dashed lines) through the thermal plasma inclusion, as discussed in the main text. Observers at positions (y) and (z) would observe the Faraday depth interfaces and and depolarized regions depicted at bottom across the axis defined by the associated lines of sight. An observer at position (x) would see little associated Faraday depth or depolarization structure. Note that the structures depicted here are not intended to be shown to scale.}
\label{fig:AdvectedScenario}
\end{figure}

As discussed in Section \ref{sec-reqmass}, both observations and simulations demonstrate that buoyant radio lobes are capable of lifting thermal ISM from their hosts and mixing it inhomogeneously throughout their volume. Indeed, this is our favored scenario for the origin of the thermal plasma associated with the low-$p$ patches. We now propose the following toy model (depicted in Figure \ref{fig:AdvectedScenario}) to account for the associated Faraday depth structure within this framework: Blobs of thermal plasma laced with $\sim\mu$G strength magnetic fields (e.g. \citealp{MS1996,OP2017}) are lifted from the host galaxy ISM or surrounding ICM and entrained in vortical flows set up in the buoyant radio lobe. The plasma is dragged into shells or filaments (e.g. see \citealp{HS2017,Weinberger2017}), shearing and stretching the embedded magnetic field into loops. From a range of viewing angles (e.g. from positions `y' and `z' indicated in Figure \ref{fig:AdvectedScenario}), lines-of-sight intersect these shell walls/filaments tangentially over considerable path lengths, generating a substantial associated Faraday depth (see below). Across the intersecting tangent plane, we see reversals in the LOS-projected magnetic field direction, and thus the sign of the peak Faraday depth. At the same time, when viewed from a different location, the same plasma filament may produce little in the way of observable Faraday rotation or depolarization (e.g. from position `x'  in Figure \ref{fig:AdvectedScenario}). In this way, regions of the lobe exhibiting enhanced Faraday depths and depolarization will be limited to sight-lines that simultaneously contain a substantial column density of advected thermal plasma, and a substantial LOS-projected magnetic field component. These two factors will tend to be positively correlated by the vortical flow patterns,  leading naturally to the generation of spatially-limited regions in which the peak Faraday depth can be substantially larger than the immediate surroundings.

To calculate the Faraday depths plausibly associated with these structures, we take \citealt{Seta2013}'s estimate of the electron density of thermal material in the lobe for $f=0.004$ (see Section \ref{sec-tpcontent}), $B_\text{av}$ from Section \ref{sec-magfield}, and the length of a typical low-$p$ patch (Section \ref{sec-sizes}) as a proxy for the path length intercepted along the wall of a thermal plasma shell, and find that Faraday depths of up to $\sim$100s rad m$^{-2}$ might be generated. With such high Faraday depths, mixing of the relativistic and thermal plasma, or inhomogeneities in the advected medium, will naturally produce large differential Faraday depths and dispersions, thereby accounting for the observed depolarization and the associated complexity of the frequency-dependent polarized signal.\\

This basic scenario has several advantages over those previously considered. Generally:

\begin{enumerate}
\item Advected filaments/shells are naturally produced in simulations on roughly the correct timescales and length-scales (e.g. see Fig. 9 of \citealt{Weinberger2017})
\item Vortical flows will naturally shear the advected magnetized thermal material into structures possessing a comparatively large characteristic scale length, and thus magnetic coherence length. As a result, the required local levels of enhancement in $n_e$ and/or $B$ over the lobe averages (discussed in Section \ref{sec-FDstruct}) are considerably relaxed.
\item Presumably, there will be no special viewing angles required to observe the interfaces --- a wide range of lines of sight will intersect the putative thermal plasma shells tangentially, producing the effect.
\item Spatial correlations between the low-$p$ patches, $\phi_{\text{peak}}$, the sky-projected magnetic field, and even Stokes $I$ filaments will all arise naturally, since each observable is generated throughout the bulk of the lobe in a global flow. In particular, the position of maximum depolarization and the $\phi_\text{peak}$ interfaces should be roughly coincident, since they both arise from looking along the wall/axis of a shell/filament.
\end{enumerate}

Nevertheless, the true viability of this scenario is not simple to assess --- it will need to be tested with simulations capable of generating high-resolution, detailed maps of Faraday depth structure in buoyant radio bubbles. At the same time, the ubiquity of these features in the radio lobe population will need to be established with broadband observations of large samples of spatially-resolved sources. Despite these caveats, we consider this basic scenario to be the most promising explanation for our observational results taken as a whole. 

\subsection{Concluding remarks}\label{sec:Conclude}

The frequency-dependent polarization behavior associated with the low-$p$ patches in Fornax A is both pronounced and remarkable. So far as we are aware, such behavior has not been observed towards radio lobes before. This may be because as an old, large, well-resolved radio source, which is neither FR I nor FR II, and which has evolved relatively passively over an extended period on the outskirts of a poor cluster, Fornax A is itself a unique object in our skies. Alternatively, it may be that other radio lobes do show such structure, but that this has formerly gone undetected. Previous spatially-resolved studies of the Faraday effect towards radio lobes have often claimed that the observed polarization angle has a linear dependence on $\lambda^2$, and that where depolarization in $p(\lambda^2)$ is observed, it can be explained by the action of a simple turbulent magnetized plasma lying in the foreground. However, these studies generally rely on interpolating polarization behavior between a handful of narrow, widely-spaced observing bands. A great deal of subtle Faraday structure can be missed in this way --- it is much like attempting to reconstruct a complex, extended object by applying aperture synthesis imaging to data from an array consisting of only a few baselines (e.g. see \citealp{BdB2005,Anderson2016b}).

Indeed, it would be surprising if this latter alternative were \emph{not} the case for well-evolved radio lobes. A large body of 2D, 2.5D (i.e. axisymmetric) and 3D hydrodynamic, magnetohydrodynamic, and relativistic magnetohydrodynamic simulations of the evolution and impact of radio lobes in various cosmic environments now exists (e.g. \citealp{Bicknell1990,BK2001,KA2007,HE2011,HK2014,EHK2016,YR2016,HS2017,Weinberger2017}, and refs. in each). These generally show that in the later stages of their evolution, radio lobes are subject to bulk vortical motions and surface instabilities, which generally reorganize and amplify frozen-in magnetic field lines, and which can advect and mix with thermal plasma from the environment. The final magneto-ionic structure of the radio lobes is generally complicated, and we can reasonably expect that complicated frequency-dependent interference effects in the polarized signal from these regions will be the norm. Perhaps these signatures have previously been masked by Faraday rotation in the deep cluster potentials that observations have typically focused on, or by some combination of poor frequency coverage and sampling, insufficient spatial resolution, intrinsic Faraday depolarization in the observing band, or insufficient accuracy attainable in polarization calibration --- these are all issues of practical importance, but separate to that of whether or not the polarization signatures are generated by the lobes in the first place. Thus, it may be that re-observing large and powerful radio sources over broad, densely-sampled bands could yield new observational insights into the physics of radio lobes. We are engaged in several current (e.g. \citealp{Kaczmarek2018}) and planned projects to do just this.

To our knowledge, existing simulations lack the volumetric resolution to predict the spectropolarimetric signatures of realistic structures generated by instabilities at the lobe-external medium interface, and perhaps the material found deeper inside. Future work along these lines will be immensely valuable for comparison against broadband observations, which may provide a powerful new capability for tracing magnetic field structure and flow patterns in radio lobes.

\section{Summary and conclusions}\label{conclusion}

We have examined the radio polarization properties of Fornax A over 1.28--3.1 GHz using RM synthesis and ($q$,$u$)-fitting. In this work (the first of two associated papers), our primary goal was to constrain the physical origin of the prominent low-$p$ patches in the lobes. Our principal findings are that:

\begin{enumerate}[i)]
\item The low-$p$ patches are in general spatially-resolved, centered around enhancements in the value of $|\phi_{\text{peak}}|$ of typically up to $\sim$50 rad m$^{-2}$ (sometimes greater), and around `interfaces' across which the Faraday depth often changes sign and jumps by $\sim50$--$100$ rad m$^{-2}$. We argued that this implies the operation of a physical process that spatially couples depolarized Faraday structures to changes in the regular line of sight magnetic field.
\item Strong and often oscillatory (as a function of $\lambda^2$) depolarization is observed in and around the low-$p$ patches. Thus, the polarized emission is not consistent with having experienced pure Faraday rotation by a magnetized foreground plasma, nor frequency-independent depolarization. Our ($q$,$u$) fitting analysis generally requires various combinations of multiple emission components, with both internal and external Faraday depolarization acting upon them, to describe the net spectropolarimetric behavior along most sight-lines. Thus, the magnetoionic structure of the Faraday-active medium is highly complicated. 
\item The low-$p$ patches sit near regions where the sky-projected magnetic field orientation possesses turbulent, small-scale structure relative to other locations in the lobe. There are also indications of a relationship between the position of low-$p$ patches and that of bright Stokes $I$ filaments, but the nature of this relationship is less clear-cut. The fact that fractional polarization in the former regions is frequency-dependent, and analysis of the gradient in the sky-projected magnetic field orientation, rules out `crossed' magnetic field lines in such regions as the direct cause of the low polarization values. Instead, the low-$p$ patches and Faraday depth interfaces appear to trace junctures in the magnetic field structure of the lobes, and by extension, bulk flows in the lobe medium.
\item We have considered several possible causes of the low-$p$ patches and the associated Faraday depth structure. We rule out, or otherwise disfavor, scenarios involving material transported into the lobes by the radio jet or external transiting galaxies, foreground material such as filaments in the ICM, and Rayleigh-Taylor or Kelvin-Helmholtz instabilities acting at the lobe-ICM surface.
\item We find that the low-$p$ patches and associated Faraday depth enhancements are associated with a minimum of $\mathcal{O}(10^9)$ $M_\odot$ of magnetized thermal plasma residing in the lobes, most likely advected from the host galaxy ISM through buoyant uplift. We have proposed a toy model that links this material to the low-$p$ patches, to the observed enhancements in peak Faraday depth, and to the sign-reversals in peak Faraday depth that occur in these locations. In short: Vortical motion in the rising lobes shear the uplifted thermal plasma into shells or filaments, with frozen-in magnetic fields being sheared into elongated loops within. Lines-of-sight intersect these shell walls/filaments tangentially over considerable path lengths, producing associated Faraday depths of perhaps up to 100s rad m$^{-2}$. The LOS-projected magnetic field reverses across the tangent plane, generating the sign reversals in peak Faraday depth. Inhomogeneities in the magnetoionic structure of this thermal material, or a co-located suffusion of synchrotron-emitting plasma, result in large Faraday dispersions and/or differential Faraday depths, generating the associated depolarization.
\item To our knowledge, and regardless of its cause, such pronounced complex frequency-dependent spectropolarimetric behavior has not yet been observed in other lobed objects. This may be because Fornax A is somewhat unique in terms of its object type, age, environment, size, or brightness, but it may also be due to previous observations lacking dense and broad $\lambda^2$ coverage. Re-observation of previously-observed `classic' radio galaxies using broad, densely-sampled observing bands may uncover similar structures. 
\item Broadband polarimetry may provide a unique and powerful new means of tracing and characterizing flow structures in radio lobes and mixing at the lobe-ICM interface. Such observations would be particularly useful for comparison with modern simulations of radio lobes in various environments. 

\end{enumerate}

\noindent In paper 2, we will present further analysis of the spectral behavior of the Fornax A lobes, over their full area, in both total intensity and linear polarization. We are also currently pursuing broadband polarimetric observations of a larger sample of lobed radio sources located in different environments to further explore the scenarios described in this work.

\section{Acknowledgements}\label{Acknowledgements}

\noindent We are indebted to the referee Jean Eilek for providing constructive feedback and criticisms of our work --- several key aspects of our discussion and conclusions have changed significantly as a result. B.~M.~G. and C.~S.~A. acknowledge the support of the Australian Research Council (ARC) through grant FL100100114. B.~M.~G. acknowledges the support of the Natural Sciences and Engineering Research Council of Canada (NSERC) through grant RGPIN-2015-05948, and of the Canada Research Chairs program. The Dunlap Institute is funded through an endowment established by the David Dunlap family and the University of Toronto. The Australia Telescope Compact Array is part of the Australia Telescope National Facility which is funded by the Commonwealth of Australia for operation as a National Facility managed by CSIRO.

\include{journals}
\bibliographystyle{apj}
\bibliography{bibliography}

\appendix

\section{Ancillary images}\label{sec-appendA}

\thispagestyle{empty}
\begin{figure*}
\centering
\includegraphics[width=0.75\textwidth]{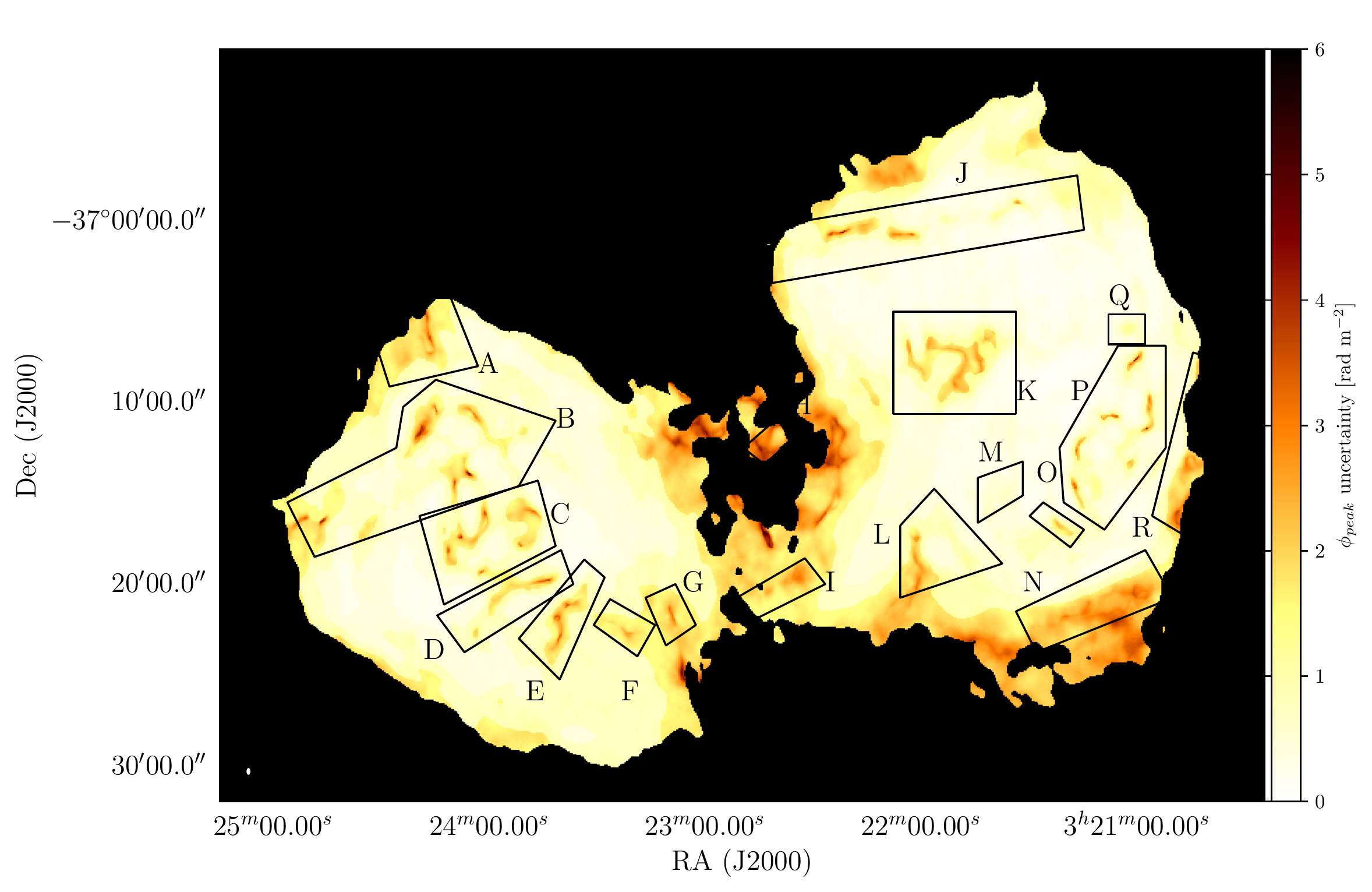}
\caption[Map of the uncertainty on the peak Faraday depth over Fornax A ($20"\times30"$ spatial resolution)]{A map of the uncertainty on the peak Faraday depth ($20"\times30"$ spatial resolution). The boxes A--R enclose the same regions as those in Figure \ref{fig:FornADepolRegions}. The white boxed ellipse in the lower left indicates the synthesized beam size. }
\label{fig:FornAFDErrRegions}
\end{figure*}

\thispagestyle{empty}
\begin{sidewaysfigure}
\centering
\includegraphics[width=1.0\textwidth,origin=c]{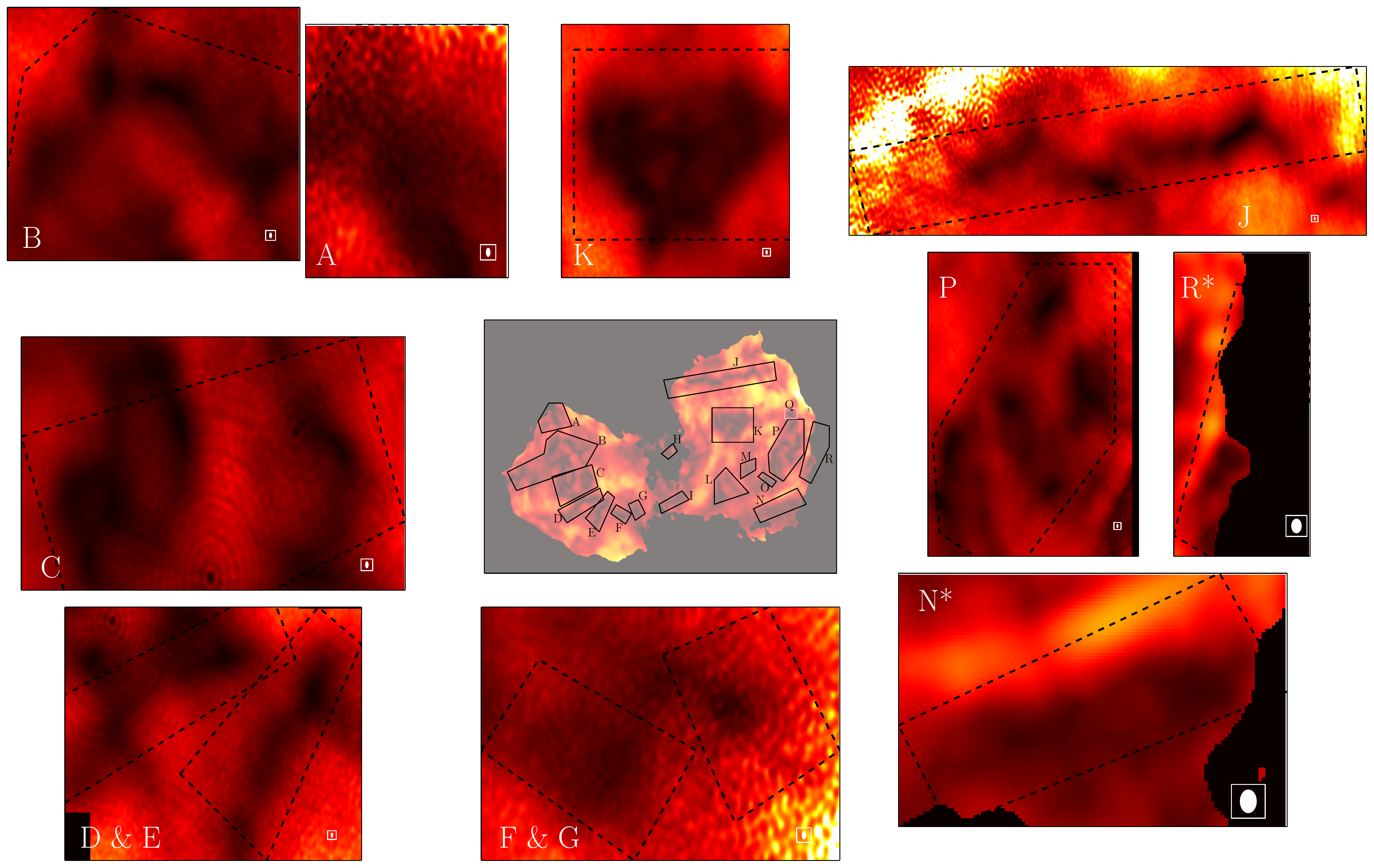}
\caption[Map of fractional polarization over Fornax A ($4"\times8"$ spatial resolution)]{\emph{Central panel:} Same as Figure \ref{fig:FornADepolRegions}. \emph{Other panels:} Maps of the magnitude of the RM synthesis $+$ \textsc{rmclean}-derived FDS (i.e. band-averaged fractional polarization) of Fornax A in selected regions at $4"\times8"$ spatial resolution in RA and Decl. respectively (see Section \ref{sec:overviewpol}). The black dashed lines delineating the regions and the letters labeling those regions match those shown in the $20"\times30"$ image displayed in the central panel. The synthesized beam size is indicated with a white ellipse in the lower right-hand corner of each panel. Regions with `starred' letter are shown at $20"\times30"$, because the S/N was too low in the $4"\times8"$ datacubes to perform RM synthesis. Note that at $4"\times8"$ resolution, the low-$p$ patches described in Section \ref{sec:overviewpol} are generally resolved by a large number of beam-widths. Note also the similarity in morphology and extent of the low-$p$ patches as they appear at both $4"\times8"$ and $20"\times30"$ spatial resolution, despite the radically different $uv$ tapering scheme adopted in each case.}
\label{fig:FornAFDZoomRegionsDP_paper}
\end{sidewaysfigure}

\thispagestyle{empty}
\begin{sidewaysfigure}
\centering
\includegraphics[width=1.0\textwidth,origin=c]{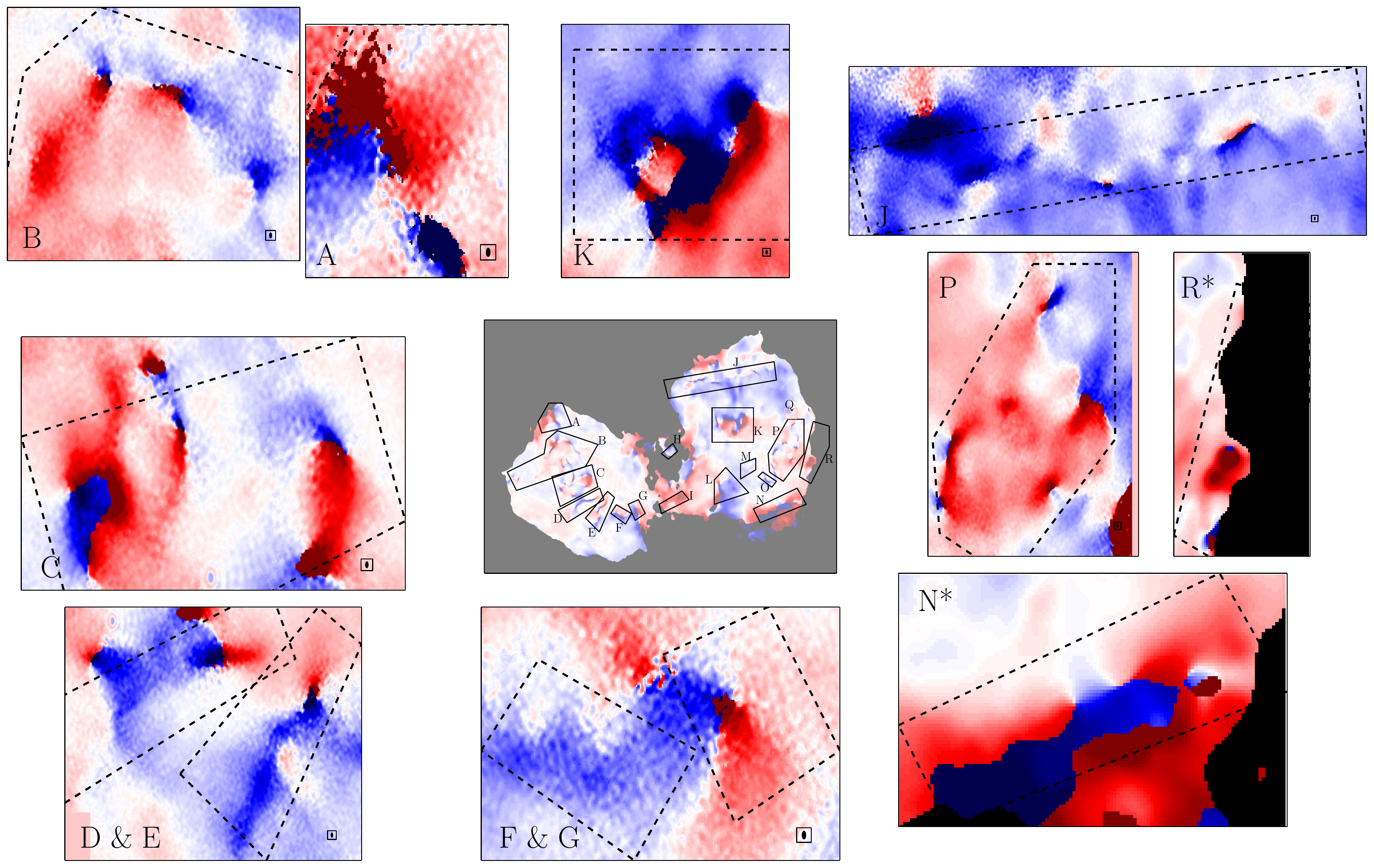}
\caption[Map of peak Faraday depth over Fornax A ($4"\times8"$ spatial resolution)]{\emph{Central panel:} Same as Figure \ref{fig:FornAFDRegions}. \emph{Other panels:} Maps of the peak Faraday depth in selected regions at $4"\times8"$ spatial resolution in RA and Decl. respectively (see Section \ref{sec:overviewpol}). The black dashed lines delineating the regions and the letters labeling those regions match those shown in the $20"\times30"$ image displayed in the central panel. The synthesized beam size is indicated with a black ellipse in the lower right-hand corner of each panel. Regions with `starred' letter are shown at $20"\times30"$, because the S/N was too low in the $4"\times8"$ datacubes to perform RM synthesis. Note that the Faraday depth structures described in Section \ref{sec:spatialFD} are now generally resolved by a large number of beam-widths. Also note the general similarity in Faraday depth structure between both the $4"\times8"$ and $20"\times30"$ maps, despite the radically different $uv$ tapering scheme adopted in each case.}
\label{fig:FornAFDZoomRegionsFD_paper}
\end{sidewaysfigure}

\end{document}